\documentclass[Phys, submission]{SciPost}

\hypersetup{breaklinks=true,colorlinks,linktocpage,citecolor=blue,linkcolor=red,urlcolor=blue}
\usepackage[capitalize]{cleveref}
\crefformat{section}{Sec.~#2#1#3}

\usepackage{subcaption}
\usepackage{amsmath}
\usepackage{mcite}

\graphicspath{{figures/}}

\numberwithin{equation}{section}

\newcommand{\dr}{{\mathrm{d}}}
\newcommand{\er}{{\mathrm{e}}}
\newcommand{\ir}{{\mathrm{i}}}
\newcommand{\frm}{{\mathrm{f}}}
\newcommand{\srm}{\mathrm{s}}

\newcommand{\Crm}{\mathrm{C}}
\newcommand{\Drm}{{\mathrm{D}}}

\newcommand{\Hrm}{{\mathrm{H}}}

\newcommand{\Lrm}{{\mathrm{L}}}
\newcommand{\Rrm}{{\mathrm{R}}}

\newcommand{\Trm}{{\mathrm{T}}}
\newcommand{\orm}{{\mathrm{o}}}
\newcommand{\erm}{{\mathrm{e}}}
\newcommand{\Erm}{{\mathrm{E}}}

\newcommand{\Ac}{\mathcal{A}}

\newcommand{\Dc}{\mathcal{D}}

\newcommand{\Gc}{\mathcal{G}}

\newcommand{\Oc}{\mathcal{O}}

\newcommand{\Wc}{\mathcal{W}}

\newcommand{\fv}{{\mathbf{f}}}

\newcommand{\rv}{{\mathbf{r}}}
\newcommand{\pv}{{\mathbf{p}}}
\newcommand{\qv}{{\mathbf{q}}}
\newcommand{\uv}{{\mathbf{u}}}

\newcommand{\xv}{{\mathbf{x}}}

\newcommand{\Dv}{{\mathbf{D}}}
\newcommand{\Ev}{{\mathbf{E}}}

\newcommand{\Jv}{{\mathbf{J}}}

\newcommand{\Avc}{\boldsymbol{\mathcal{A}}}

\newcommand{\Deltav}{{\boldsymbol{\Delta}}}
\newcommand{\Psiv}{\boldsymbol{\Psi}}

\newcommand{\nablav}{{\boldsymbol{\nabla}}}

\newcommand{\Bsf}{\mathsf{B}}
\newcommand{\Esfv}{\boldsymbol{\mathsf{E}}}

\newcommand{\cs}{c_{s}}
\newcommand{\cst}{\tilde{c}_{s}}

\newcommand{\Kt}{\tilde{K}}
\newcommand{\Qt}{\tilde{Q}}

\newcommand{\Act}{\tilde{\mathcal{A}}}

\newcommand{\sigt}{\tilde{\sigma}}

\newcommand{\braket}[1]{\ensuremath{\langle#1\rangle}}

\newcommand{\Uone}{{\mathrm{U}(1)}}
\newcommand{\Urm}{\mathrm{U}}
\newcommand{\SO}{\mathrm{SO}}

\DeclareMathOperator{\tr}{tr}
\DeclareMathOperator{\Tr}{Tr}
\DeclareMathOperator{\re}{Re}
\DeclareMathOperator{\im}{Im}

\newcommand{\nnl}{\nonumber \\}

\newcommand{\beq}{\begin{equation}}
\newcommand{\eeq}{\end{equation}}
\newcommand{\veps}{\varepsilon}
\usepackage{mathrsfs} 

\usepackage[USenglish]{babel}
\usepackage{microtype}

\begin{document}

\begin{center}{\Large \textbf{
Hall   viscosity and conductivity of  two-dimensional chiral  superconductors
}}\end{center}

\begin{center}
F\'{e}lix Rose\textsuperscript{1,2,4*},
Omri Golan\textsuperscript{3},
Sergej Moroz\textsuperscript{1,4}
\end{center}

\begin{center}
{\bf 1} Physik-Department, Technische Universität München, 85748 Garching, Germany
\\
{\bf 2} Max-Planck-Institut für Quantenoptik, 85748 Garching, Germany
\\
{\bf 3} Department of Condensed Matter Physics, Weizmann Institute of Science, \\
Rehovot 76100, Israel
\\
{\bf 4} Munich Center for Quantum Science and Technology (MCQST), \\
80799 München, Germany
\\
* \href{mailto:felix.rose@m4x.org}{felix.rose@m4x.org}
\end{center}

\begin{center}
April 6, 2020
\end{center}

% For convenience during refereeing: line numbers
%\linenumbers

\section*{Abstract}
{\bf
We compute the Hall  viscosity and conductivity of non-relativistic two-di\-mensional chiral superconductors, where fermions pair due to a short-range attractive potential, e.g. $\boldsymbol{p+\mathrm{i}p}$ pairing, \textit{and} interact via a long-range repulsive
Coulomb force. For a logarithmic Coulomb potential,  the Hall viscosity tensor contains a contribution that is singular at low momentum, which encodes corrections to pressure induced by an external shear strain. Due to this contribution, the Hall viscosity cannot be extracted from the Hall conductivity in spite of Galilean symmetry. For mixed-dimensional chiral superconductors, where the Coulomb potential decays as inverse distance, we find an intermediate behavior between intrinsic two-dimensional superconductors and superfluids. These results are obtained by means of both effective and microscopic field theory.

}

\vspace{10pt}
\noindent\rule{\textwidth}{1pt}
\tableofcontents\thispagestyle{fancy}
\noindent\rule{\textwidth}{1pt}
\vspace{10pt}

\section{Introduction}
\label{sec:intro}
Two-dimensional chiral pairing, with its fully gapped  Fermi surface and Cooper pairs that coherently carry a finite angular momentum, is a theoretical paradigm of a quantum topological phase of matter \cite{Read2000, Alicea2012, volovikbook} 
which is nowadays under intense experimental investigation. It was discovered recently that the chiral A-phase of superfluid $^3$He becomes stable at zero temperature under nanoscale confinement \cite{Levitin2013, *Shook2019}. New experimental signatures of topological superconductivity \cite{kallin2016chiral} were also reported in thin superconducting films \cite{Xu2015, *Menard2017}. A recent experiment with the 5/2 quantum Hall state \cite{banerjee2018observation}, which theoretically is believed to be some chirally paired superconductor of composite fermions, reignited the long-term debate about the nature of topological order of this state.

Fermionic chiral paired states exhibit non-dissipative Hall responses because the chiral order parameter breaks time-reversal $T$ and parity $P$ symmetries spontaneously. The well-known Hall conductivity tensor $\sigma^{ij}_\Hrm(\omega, \mathbf{q})$ quantifies the response of the $\Urm(1)$ current $\mathbf{J}(\omega, \mathbf{q})$ to a monochromatic electric field $\mathbf{E}(\omega, \mathbf{q})$. The Hall conductivity in a two-dimensional neutral chiral superfluid at zero temperature  was computed in \cite{goryo1998abelian, *horovitz2002spontaneous, *PhysRevB.77.174513, Lutchyn2008, Hoyos2013}. 
 In addition to the Hall conductivity, a clean two-dimensional system is characterized by a supplementary non-dissipative Hall response, the Hall (or odd) viscosity tensor $\eta^{ijkl}_\orm(\omega, \mathbf{q})$, that fixes the (odd under time-reversal) response of the stress tensor to the strain rate \cite{Avron1995, Avron1997}, see \cite{Hoyos2014} for a review. Recently, observable signatures of the Hall viscosity have been vigorously studied in classical and quantum fluids  both theoretically \cite{PhysRevB.94.125427,PhysRevLett.119.226602,PhysRevLett.118.226601,holder2019unified,PhysRevB.100.115421,PhysRevE.90.063005,PhysRevB.96.174524,PhysRevFluids.2.094101,banerjee2017odd,PhysRevLett.122.128001, abanov2019hydrodynamics, PhysRevLett.123.026804, narozhny2019magnetohydrodynamics, PhysRevB.100.115401} and experimentally \cite{soni2018free, Berdyugineaau0685}. In a two-dimensional isotropic system that is invariant under the combined $PT$ symmetry the odd viscosity tensor reduces to two independent components \cite{Golan2019}, in this paper to be denoted $\eta_\orm^{(1)}(\omega, \mathbf{q}^2)$ and $\eta_\orm^{(2)}(\omega, \mathbf{q}^2)$, respectively. If the Hall viscosity tensor is regular in the limit $\mathbf{q}=0$, only the component $\eta_\orm^{(1)}$ survives as $\mathbf{q}\to 0$ \cite{Avron1995, Avron1997}.  For gapped quantum fluids the ratio of the Hall viscosity to the particle number density $n_0$  was argued to be quantized in the units of $\hbar$ as follows \cite{Read2009, Read2011}
\beq
\frac{\eta_\orm^{(1)}}{n_0}=\frac 1 2 s,
\eeq
where $s$ is a rational number that is equal to the average angular momentum per particle. In a neutral $l$-wave chiral superfluid, though gapless, the same relation holds. Here $s=\pm l/2$ with the sign fixed by the chirality of the condensate. As a result, at $\mathbf{q}=0$ the Hall viscosity coefficient $\eta_\orm^{(1)}$ does not depend on the topology of the fermionic ground state and thus cannot be used as a diagnostics of topological superconductivity that is characterized by protected chiral Majorana edge modes. It was shown however in \cite{Golan2019}  that the $\mathbf{q}^2$ dependence of the Hall viscosity tensor contains information about the chiral central charge of the boundary theory, which is determined by the topology of the fermionic ground state.

In single-component Galilean-invariant fluids and solids with a particle number symmetry, the momentum density is proportional to the particle number current, resulting in a Ward identity that ties together the viscosity and conductivity tensors\cite{taylor2010viscosity, Hoyos2012, Bradlyn2012, Geracie2015}. For the two-dimensional chiral superfluid the relation acquires a simple form in the uniform limit $q=|\qv| \to 0$
\begin{equation} \label{Wardsup}
\eta^{(1)}_{\orm}(\omega) = -\frac{m^2 \omega^2}{2} \partial_{q}^2 \sigma_{\Hrm} (\omega,\qv) \big|_{\qv=0}.
\end{equation}
As a result, in this system one can extract the AC Hall viscosity  $\eta^{(1)}_{\orm}(\omega)$ only from the knowledge of the Hall conductivity at small momentum.

The main aim of this paper is to compute the Hall conductivity and viscosity of a two-dimensional chiral superconductor, where chirally paired fermions couple to a fluctuating electromagnetic field. Two different types of such superconductors can be considered. In a mixed-dimensional  superconductor, fermions are restricted to a surface, but the electromagnetic field extends in full three-dimensional space.  In such a superconductor plasma oscillations are gapless and charges and vortices exhibit long-range interactions \cite{Marino2018}. Electromagnetic response in the mixed-dimensional chiral superconductor was computed in \cite{Lutchyn2008}.\footnote{It was also argued recently in \cite{boyack2019electromagnetic} that the order parameter and Coulomb fluctuations do not contribute to the Meissner effect in the chiral $p$-wave superconductor.}
 Alternatively, one may consider an intrinsic two-dimensional chiral superconductor, where akin to fermions the electromagnetic field is restricted to the surface. As a result, plasma oscillations are gapped. While this case might seem somewhat contrived, there are two physical motivations to investigate it: (i) In quantum Hall fluids composite fermions couple to a fluctuating emergent $u(1)$ gauge field that is defined in $(2+1)$-dimensional spacetime \cite{Halperin1993, Son2015}. As a result, chiral paired states of composite fermions are intrinsic two-dimensional chiral superconductors. (ii) Sufficiently small planar Josephson junction arrays can realize an intrinsic two-dimensional conventional superconductor \cite{Mooij1990, *Fazio1991, *Diamantini1996, *Baturina2013}. This naturally suggests that a planar Josephson junction array of chiral superconducting islands can give rise to an intrinsic two-dimensional chiral superconductor. 

In this work we consider non-relativistic fermions, and  accordingly, approximate the interaction mediated by the gauge field by an instantaneous Coulomb potential. As a result the problem we study is Galilean invariant. Within this setting we develop a unified theoretical framework for intrinsic and mixed-dimensional chiral superconductors, which we use to extract electromagnetic and gravitational linear responses and investigate the conductivity-viscosity relations following from Galilean symmetry.

Our main results are summarized in \cref{sec:sum}. The rest of the paper is structured as follows: In \cref{sec:symAndWard} we present a general discussion of electromagnetic and geometric linear responses and the Ward identities which tie them together in Galilean-invariant systems. In \cref{sec:EFT} we develop a low-energy effective field theory of different types of chiral superconductors and present a streamlined calculation of the Hall conductivity and Hall viscosities. In \cref{sec:microTheory} we reproduce our results for the Hall responses directly from a canonical microscopic fermionic model of a two-dimensional chiral superconductor. We conclude our work with \cref{sec:conc}, where we provide an outlook for future research.

 \section{Main results and physical picture} \label{sec:sum}
 
 In this section, we  summarize and explain our main results for the intrinsic two-dimensional chiral superconductor. The derivation of these results and their extension to an arbitrary long-range Coulomb interaction can be found in the following sections.
 
 \subsection{Setup}
 
 Before presenting our results, we emphasize that the Hall conductivity and viscosity computed in this paper are defined as linear responses to \textit{external sources}, as opposed to total fields. As is well known, the total electric field $\mathbf{E}-\boldsymbol{\nabla}\chi$ contains the externally applied field $\mathbf{E}$ and the internal contribution $-\boldsymbol{\nabla}\chi$, where $\chi$ is the electric potential generated by charges in the system. The conductivity we compute is then defined as the response of current to an applied $\mathbf{E}$, and accounts, in particular, for the potential $\chi$ induced due to the applied electric field. This conductivity is physically relevant when an external field $\mathbf{E}$ is applied in the bulk of the system, away from boundaries. 
 
 Less appreciated is the analogous decomposition of the strain-rate in the context of the viscosity calculation. The total strain-rate tensor $\dot{u}_{ij}+\partial_{(i}v_{j)}$ contains the externally applied strain $u_{ij}$, which corresponds to a background spatial metric $g_{ij}=\delta_{ij}+2u_{ij}$, as well as the internal strain-rate $\partial_{(i}v_{j)}$, where $v_{j}$ is the velocity of particles in the system  and the parenthesis denote symmetrization~\cite{Hoyos2013}. The viscosity we compute in this paper is defined as the response of stress to an applied $\dot{u}_{ij}$, and accounts, in particular, for the velocity $v_{i}$ induced due to the applied external strain $u_{ij}$. This viscosity is physically relevant when an external strain rate $\dot{u}_{ij}$ is applied in the bulk of the superconductor away from boundaries, as in \cref{fig:presIndVis}. This should be contrasted with standard hydrodynamic scenarios, where $u_{ij}=0$ and the system is perturbed via boundary conditions \cite{Lucas2014}.

  In a crystalline superconductor the strain $u_{ij}$  naturally describes the crystal structure of the ion background \cite{Hughes2013}, and the odd viscosity we compute modifies the dispersion of the corresponding phonon excitations, à la Ref.~\cite{Barkeshli2012}. 
  It will be useful to analyze our  results in terms of the ion displacement field  $\boldsymbol{\xi}\left(\mathbf{x}\right)=\delta \mathbf{x}$, assumed to be in-plane for simplicity. Then  $u_{ij}=\partial_{(i}\xi_{j)}$, and the total strain rate is given by $\partial_{(i}\dot{\xi}_{j)}+\partial_{(i}v_{j)}$, where the ion velocity $\dot{\xi}_{i}$ and electron velocity $v_{i}$ enter symmetrically.

\subsection{Hall conductivity and viscosity}

\begin{figure}[t]
\centering
    \begin{subfigure}[m]{0.45\textwidth}
    \includegraphics[width=\textwidth]{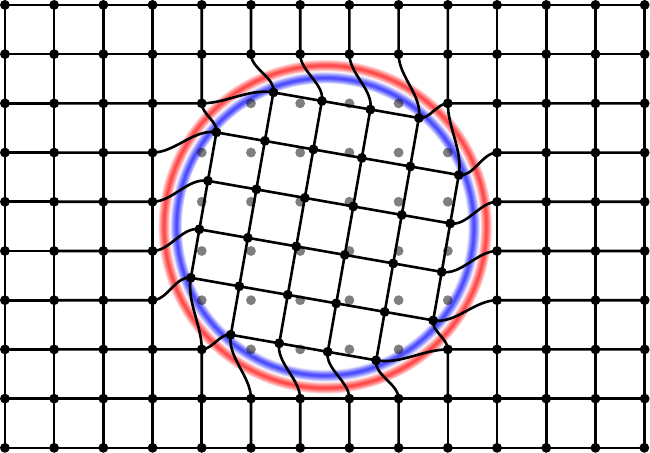}
        \caption{Superfluid.}
        \label{fig:rotvisSF}
    \end{subfigure}
\hspace{0.05\textwidth}
    \begin{subfigure}[m]{0.45\textwidth}
    \includegraphics[width=\textwidth]{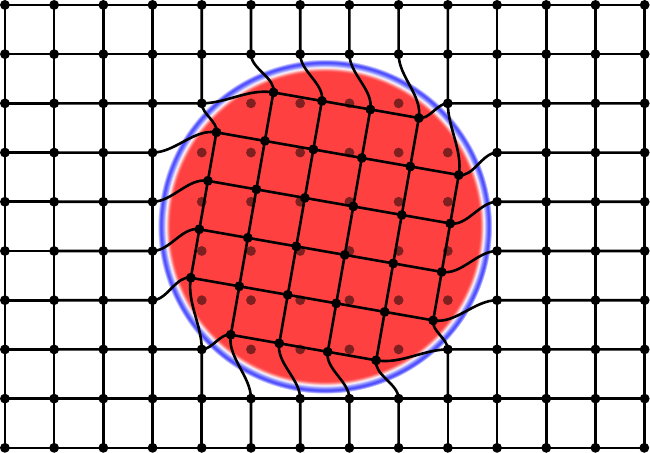}
        \caption{Superconductor.}
        \label{fig:rotvisSC}
    \end{subfigure}
\caption{Pressure profile induced through the Hall viscosity coefficient $\eta^{(2)}_\orm$  by an applied external strain in a chiral superfluid (left) and superconductor (right). A disk of radius $R$ is rotated at a frequency $\omega$ around its center, the rotation corresponding to a radial displacement field $\xi_i = \Phi(t,r) \epsilon_{ij} x^j$ with $\Phi(t,r)$ independent of $r$ inside the disk and vanishing outside. The displacement field at a given time is represented by its effect on a black grid, with the untransformed points plotted in grey. The corresponding vorticity $\Omega = -2 \dot{\Phi} - r\partial_{r} \dot{\Phi}$ induces  through \cref{eq:pressVorrel} a change in pressure $\delta p$ shown in color, with red, blue and white corresponding to $\delta p$ positive, negative and vanishing, respectively. In the superfluid case $\eta^{(2)}_\orm$ is independent of $\qv$ at long wavelengths $c_{s}| \mathbf{q} | \ll \omega$ [see \cref{EFT:eta12cs}] and the pressure shift $\delta p \sim \nablav^{2} \Omega$ is localized at the edge of the disk. On the other hand, for a superconductor, where $\eta^{(2)}_\orm \sim 1/\qv^2$ for $c_{s}| \mathbf{q} |$, $ \omega \ll \omega_{p}$ [see \cref{eq:eta12}],  the vorticity inside of the disk is directly related to the pressure shift, $\delta p \sim \Omega$. For a given vorticity, the sign of the resulting pressure change is determined by the chirality of the superfluid/superconductor.
\label{fig:presIndVis}
}
\end{figure}

The Hall conductivity at finite frequency $\omega$ and momentum $\mathbf{q}$ that we find is given by
\beq \label{sum1}
\sigma_{\Hrm}(\omega, \mathbf{q}^2)=\frac 1 2 \epsilon_{ij} \sigma^{ij}(\omega, \mathbf{q})=\frac{s n_0}{2m^2} \frac{-\mathbf{q}^{2}}{\omega^{2}-\omega_p^2-c_{s}^{2} \mathbf{q}^{2}},
\eeq
where we introduced the plasma frequency $\omega_p=\sqrt{e^2 n_0/m}$. In contrast to the neutral chiral  superfluid, where $e=0$, the energy gap of the plasmon excitation ensures that the Hall conductivity always vanishes as $\mathbf{q}^2$ in the limit $\omega$, $\mathbf{q}\to 0$.

The Hall viscosity tensor of an isotropic $PT$-invariant system  is given by \cite{Golan2019}
\begin{equation} 
\eta_\orm(\omega,\qv) =  \eta_\orm^{(1)}(\omega,\qv^2) \sigma^{xz} +  \eta_\orm^{(2)}(\omega,\qv^2)[(q_x^2-q_y^2)\sigma^{0x}-2q_x q_y \sigma^{0z}], \label{sum2}
\end{equation} 
and is fixed by the two independent coefficients $\eta_\orm^{(1)}(\omega,\qv^2)$ and $\eta_\orm^{(2)}(\omega,\qv^2)$, where $\sigma^{ab}=\sigma^{a}\otimes \sigma^{b}-\sigma^{b}\otimes \sigma^{a}$ , $a,b=0,x,z$, are anti-symmetrized tensor products of symmetric Pauli matrices, see \cref{sec:symAndWard}. 
The physical content of \cref{sum2} is as follows. The external strain $u_{ij}=\partial_{(i}\xi_{j)}$ can be decomposed into  its trace, the compression $u=u_{i}^i=\boldsymbol{\nabla}\cdot \boldsymbol\xi$, and two traceless shears. 
The viscous stress $T^{ij}=-\eta_\orm^{ijkl}\dot{u}_{kl}$  is similarly decomposed into the correction to pressure $\delta p=T^i_i/2$, and two shear stresses. While the usual odd viscosity $\eta_\orm^{(1)}$ corresponds to a Hall response of the shear stress  to a shearing rate,  and does not involve the pressure and compression, the component  $\eta_\orm^{(2)}$ induces pressure in response to a shearing rate, and the shear stress in response to a compression rate $\dot{u}$.

For the intrinsic chiral superconductor we find
\beq 
\eta_{\mathrm{o}}^{(1)}(\omega, \mathbf{q}^2)=\frac{s n_{0}}{2}, \quad \quad \eta_{\mathrm{o}}^{(2)}(\omega, \mathbf{q}^2)=-\frac{s n_{0}}{2} \frac{1}{\mathbf{q}^2} \frac{\omega_p^2+c_{s}^{2} \mathbf{q}^2}{\omega^{2}-\omega_p^2-c_{s}^{2} \mathbf{q}^{2}}. \label{eq:eta12}
\eeq
While the first component $\eta_{\mathrm{o}}^{(1)}$ coincides with the result found for the  neutral chiral superfluid, the second component $\eta_{\mathrm{o}}^{(2)}$ does not, and
exhibits a peculiar $1/\mathbf{q}^{2}$ singularity at low momenta. Combining \cref{sum2,eq:eta12}, we see that due to this singularity, the contribution of $\eta_{\mathrm{o}}^{(2)}$ to the odd viscosity tensor $\eta_{\mathrm{o}}$ does not vanish in the uniform limit $\mathbf{q}\rightarrow0$, as one may naively expect from \cref{sum2}. In fact, the Hall viscosity tensor of the intrinsic two-dimensional superconductor is ill defined at $\mathbf{q}=0$, though the superconductor is fully gapped! A similar behavior occurs in the  neutral chiral superfluid \textit{only} when it is incompressible, and  sound waves propagate with an infinite speed,  $c_s=\infty$.\footnote{Naively, a similar $1/\mathbf{q}^2$ singularity appears in the compressible neutral chiral superfluid at $\omega=0$. However, the viscosity is a response to strain-rate, and is only meaningful for $\omega\neq0$.  }

We currently understand the singularity in  $\eta_{\mathrm{o}}^{(2)}$ as originating from the instantaneous long-range nature of the Coulomb interaction. In \cref{sec:conc} we provide an  outlook for a future study that will test this understanding.

In order to demonstrate the physical significance of  $\eta^{(2)}_\orm$ and its $1/\mathbf{q}^{2}$ singularity, we note that \cref{sum2} implies 
\begin{align}
\delta p&=\eta_\orm^{(2)}\boldsymbol{\nabla}^2 \Omega,
\label{eq:pressVorrel}
\end{align}
where $\delta p$ is the pressure relative to the ground state pressure, and $\Omega=\epsilon^{ij}\partial_i \dot{\xi}_j$ is the \textit{applied} vorticity, to be distinguished from the electron vorticity  $\epsilon^{ij}\partial_i v_j$, and $\eta_\orm^{(2)}$ is the operator obtained by Fourier transforming $\eta_{\mathrm{o}}^{(2)}(\omega, \mathbf{q}^2)$ in \cref{eq:eta12}. We see that  $\eta^{(2)}_\orm$ encodes an exotic dissipationless response of pressure to applied vorticity. As follows from \cref{eq:pressVorrel}, a region of space rotating uniformly, where $\Omega$ is constant, does not support pressure variations, unless $\eta^{(2)}_\orm \sim 1/\mathbf{q}^{2}$ as $\mathbf{q}\rightarrow0$, which is illustrated in  \cref{fig:presIndVis}.

A pressure in response to uniform vorticity is currently believed to appear only in the presence of angular  momentum non-conservation, as recently discussed in Ref.~\cite{souslov2019anisotropic}. In contrast, chiral superconductors and superfluids, studied here, break rotation symmetry only \textit{spontaneously}, and are therefore isotropic and angular momentum conserving.\footnote{Technically, \cref{eq:pressVorrel} stems from the momentum dependence of the tensor multiplying $\eta^{(2)}_\orm$ in \cref{sum2}, while in angular momentum non-conserving fluids, a pressure in response to vorticity derives from the asymmetry of $\eta_\orm^{ij,kl}$ under $k\leftrightarrow l$.} Nevertheless, we see that, due to the $1/\mathbf{q}^{2}$ singularity in $\eta^{(2)}_\orm$, chiral superconductors mimic the response  $\delta p \sim \Omega$ of angular momentum non-conserving fluids, as shown in  \cref{fig:rotvisSC}.

In addition to the relation \labelcref{eq:pressVorrel}, \cref{sum2} also implies 
\begin{align}
\nablav \times \fv&= \eta_\orm^{(2)} \boldsymbol{\nabla}^4 \dot{u},
\label{eq:pressVorrel2}
\end{align} 
where $f^i = -\partial_j T^{j}_{i}$ is the force density exerted on a test particle by the electron fluid,\footnote{The force is defined as the time derivative of the momentum density, $f_i=\dot{P}_{i}$. The momentum conservation equation $\dot{P}_i = -\partial_j T^{j}_{i}$ holds only in flat space and in the absence of external forces. Here we apply a space-dependent compression, which modifies it to $\dot{P}_i = -\partial_j T^{j}_{i}+p_{0}\partial_{i}u$, where $p_{0}$ is the ground state pressure. The momentum source $\partial_{i}u$ is longitudinal, and does not contribute to $\epsilon^{ij}\partial_{i}f_j$.}
 and  $\nablav \times \fv = \epsilon^{ij}\partial_{i}f_j$ is its curl. We see that circulating forces are generated in response to an applied, space-dependent, compression rate. The $ 1/\mathbf{q}^2$ singularity in  $\eta_\orm^{(2)}$ leads to a reduction $\nablav^4\to\nablav^2$ in \cref{eq:pressVorrel2}.

\cref{eq:pressVorrel2} describes a fully geometric chiral analog of the London diamagnetic response, $\nablav \times \Jv=- \rho_\text{L} \boldsymbol{\nabla}^2 B$,  that leads to the Meisner effect in superconductors, where $\rho_\text{L}\sim 1/\mathbf{q}^2$ as $\mathbf{q}\rightarrow0$. In fact,  in \textit{chiral} superconductors, this generalizes to     $\nablav \times \Jv=- \rho_\text{L} \boldsymbol{\nabla}^2 [B+(s/2)R]$, where $R=\partial_{i}\partial_{j}u_{ij}-\nablav^{2}u$ is the (linearized) curvature, leading to spontaneous magnetization on curved surfaces  \cite{PhysRevLett.120.217002}. Since, in a Galilean invariant system, the force  is related to the electron current by $\mathbf{f}=m\dot{\mathbf{J}}$, \cref{eq:pressVorrel2} implies an additional contribution  
\begin{align}
    \nablav \times\Jv&=(\eta_\orm^{(2)}/m) \boldsymbol{\nabla}^4 u
\end{align} due to $\eta_\orm^{(2)}$, which should be taken into account in future studies of geometrically induced magnetization.

\subsection{Galilean viscosity-conductivity relation}

 Another consequence of the $1/\qv^2$ singularity in \cref{eq:eta12} is observed in the relation between the Hall conductivity and the Hall viscosities.
 Since the theory of a two-dimensional superconductor, where electromagnetic interactions are approximated by an instantaneous Coulomb potential, is Galilean invariant, the Hall conductivity and viscosities are related  via the Ward identity 
 \begin{equation} \label{sum4}
m^2 \omega^2 \sigma_\Hrm (\omega,\qv^2) = -\qv^2 [\eta_{\orm}^{(1)}(\omega,\qv^2) -\qv^2 \eta_{\orm}^{(2)}(\omega,\qv^2)],
\end{equation}
see \cref{sec:symAndWard}. Due to the $1/\mathbf{q}^{2}$ singularity of $\eta_{\mathrm{o}}^{(2)}$, 
 the relation \labelcref{Wardsup} for the uniform Hall viscosity $\eta_\orm^{(1)}(\omega)=\eta_\orm^{(1)}(\omega,\mathbf{q}=\mathbf{0})$  does not hold, and the Hall viscosity tensor $\eta_\orm(\omega,\mathbf{q})$ cannot be extracted from the Hall conductivity alone, even in the uniform limit $\mathbf{q}\rightarrow0$. 
 
Our results for the intrinsic two-dimensional superconductor demonstrate that the two independent components of the Hall viscosity, and  $\eta_\orm^{(1)}$ in particular, cannot generally  be extracted from the Hall conductivity. This does not contradict, but should be contrasted, with  recent theoretical and experimental work extracting $\eta_\orm^{(1)}$ from the Hall conductivity or from current profiles   \cite{Hoyos2012, PhysRevB.94.125427, PhysRevB.100.115421, PhysRevLett.118.226601, PhysRevLett.119.226602, holder2019unified, Berdyugineaau0685}.

As opposed to the intrinsic two-dimensional chiral superconductor, we find that  the Ward identity still takes the simple form \labelcref{Wardsup} in the mixed-dimensional chiral superconductor.

\section{Symmetries and transport coefficients}
\label{sec:symAndWard}

Consider  a quantum field theory 
at zero temperature in two spatial dimensions. The response of the system coupled to an external $\Uone$ gauge field $A_\mu$ and a spatial metric $g_{ij}$ is encoded into the induced action $\Wc[A_\mu,g_{ij}]$. The induced action can be obtained from the microscopic action $S$ by integrating out the fluctuating degrees of freedom,
\begin{equation}
\er^{\ir \Wc[A,g]} = \int \Dc[\cdots] \er^{\ir S[\cdots;A,g]}
\end{equation}
where the dots stand for the fluctuating fields.

In this Section, we focus on the symmetry properties of the induced action and thus do not specify the precise form of the action $S$, nor the dynamic degrees of freedom we consider, which depend on the scale at which we seek to describe the system. 
Following ~\cite{Son2006,Bradlyn2012,Hoyos2014,Geracie2015} we present results in real time\footnote{However, getting a proper derivation of some expressions, e.g. \cref{eq:relWsig,eq:relWeta} or \cref{eq:WardMaster} is better done using the imaginary time formalism.} and $x=(t,\xv)$ stands for a $(2+1)$-dimensional space-time variable with Fourier transform $q=(\omega,\qv)$. We use $\int_x$ as a shorthand for $\int \dr t\, \dr^2 \xv$.

\subsection{Current, stress tensor and linear response from induced action}

In this paper the $\Uone$ gauge field $A_\mu$ and the metric $g_{ij}$ are not dynamical fields but rather act as external sources. By differentiating the microscopic action with respect to the sources, one gets the current densities
\begin{align}
 J^{\mu}(x)&=-\dfrac{1}{\sqrt{g(x)}} \dfrac{\delta S}{\delta A_\mu(x)}, &
 T^{ij}(x)&=\dfrac{2}{\sqrt{g(x)}}\dfrac{\delta S}{\delta g_{ij}(x)} \label{eq:Current}
\end{align}
with $g=\det(g_{ij})$.
The expectation values of the operators $J^\mu$ and $T^{ij}$ can be obtained by replacing the action $S$  in \cref{eq:CurrentExpVal} with the induced action $\Wc$, i.e., 

\begin{align}
 \braket{J^{\mu}(x)}&=-\dfrac{1}{\sqrt{g(x)}} \dfrac{\delta \Wc}{\delta A_\mu(x)}, &
 \braket{T^{ij}(x)}&=\dfrac{2}{\sqrt{g(x)}}\dfrac{\delta \Wc}{\delta g_{ij}(x)}. \label{eq:CurrentExpVal}
\end{align}

The transport coefficients of the theory relate the response of the expectation values of the current densities to an infinitesimal change of the sources. Introducing $h_{ij} = g_{ij} - \delta_{ij}$, one has to leading order in the sources
\begin{align}
\delta \braket{J^{i}(x)} ={}&  \int_{x'} \sigma^{ij}(x-x') [\partial_t  A_{j}(x')-\partial_j  A_{t}(x')] + \Oc(A^2,h),
\label{eq:defSigLinResp}
\\
\delta \braket{T^{ij}(x)} ={}&  - \dfrac{1}{2}\int_{x'} \lambda^{ijkl}(x-x') h_{kl}(x') -  \dfrac{1}{2}\int_{x'} \eta^{ijkl}(x-x') \partial_t h_{kl}(x') + \Oc(A,h^2),
\label{eq:defViscLinResp}
\end{align}
which defines the conductivity tensor $\sigma^{ij}(x)$, the elastic modulus tensor $ \lambda^{ijkl}(x)$ and the viscosity tensor $\eta^{ijkl}(x)$. 

As $\Wc$ is the generating functional of connected correlation functions of currents, the transport coefficients can be expressed in terms of its functional derivatives 
\begin{equation}
 \Wc^{(n,m) \{ \mu_a \}, \{ i_b j_b\} }[\{x_a\}, \{y_b\};A,g]  =\dfrac{\delta^{n+m}\Wc[A,g]}{\delta A_{\mu_1}(x_1)\dots \delta A_{\mu_n}(x_n) \delta g_{i_1 j_1}(y_1)\dots  \delta g_{i_m j_m}(y_m)}.
 \label{eq:defVertices}
\end{equation}
In the following we denote the vertices evaluated in flat space $h_{ij}=0$ and vanishing $\Urm(1)$ gauge field $A_\mu$ by 
$
 \Wc^{(n,m) \{ \mu_a \}, \{ i_b j_b\} }(\{x_a\}, \{y_b\})
$. Due to translation invariance the two-point vertices are diagonal in Fourier space, for instance $\Wc^{(2,0)ij}(q,q')=\Wc^{(2,0)ij}(q)\delta_{q,-q'}$.

By further differentiating \cref{eq:CurrentExpVal} and using the definitions \cref{eq:defSigLinResp,eq:defViscLinResp} one gets
\begin{align}
\Wc^{(2,0) ij} (\omega,\qv) &= \ir \omega^{+} \sigma^{ij}(\omega,\qv), \label{eq:relWsig} \\
\Wc^{(0,2) ij kl} (\omega,\qv) &=\dfrac{1}{4}\braket{T^{ij}}\delta^{kl}-\dfrac{1}{4}\lambda^{ijkl}(\omega,\qv) + \dfrac{\ir \omega^{+}}{4} \eta^{ijkl}(\omega,\qv)\label{eq:relWeta}
\end{align}
with $\braket{T^{ij}}=\braket{T^{ij}(\omega=0,\qv=0)}$ being the homogeneous expectation value of the stress tensor for vanishing sources. In \cref{eq:relWsig,eq:relWeta}, the left-hand-sides stand for the retarded correlation functions and $\omega^{+}=\omega+\ir 0^{+}$. The infinitesimal imaginary part $0^+$ enforces causality in \cref{eq:defSigLinResp,eq:defViscLinResp}.

\subsection{Tensor decompositions of conductivity and viscosity}

 Both conductivity and viscosity  must transform as tensors under $\SO(2)$ rotations. The conductivity of an isotropic system can be decomposed as\footnote{We exclude in \cref{TensorCond} an $\SO(2)$ invariant term $\propto q^{i} \epsilon^{jk}q_k+ \epsilon^{ik}q_k q^j$. Such a term leads to dissipation $J^iE_i$  with an unconstrained sign, which is in conflict with the second law of thermodynamics. %In a $PT$ symmetric system, such a term must be multiplied by a $PT$-odd parameter (such as $s$), and therefore has an unconstrained sign, in conflict with the second law of thermodynamics. 
 }
\begin{align}
 \sigma^{ij}(\omega,\qv) = (q^i q^j/\qv^2) \sigma_\Lrm (\omega,\qv^2) + (\delta^{ij} - q^i q^j/\qv^2) \sigma_\Trm (\omega,\qv^2) + \epsilon^{ij} \sigma_\Hrm (\omega,\qv^2)\label{TensorCond}
\end{align}
with $\epsilon^{xy}=+1$, $\epsilon^{ij}=-\epsilon^{ji}$ the Levi-Civita symbol, and $\sigma_\Lrm$, $\sigma_\Trm$ and $\sigma_\Hrm$ being the longitudinal, transverse and Hall conductivities, respectively. The longitudinal and transverse components constitute the symmetric part of the conductivity tensor, satisfying $\sigma^{ij}=\sigma^{ji}$, while the Hall component fixes its antisymmetric part, $\sigma^{ij}=-\sigma^{ji}$, and describes dissipationless transport of particles. The Hall conductivity $\sigma_\Hrm(\omega, \mathbf{q})$ vanishes unless time-reversal symmetry is broken.

A similar decomposition is possible for the viscosity tensor. First, we note that by construction, it must be invariant under exchange of either the first or second pair of indices,
$\eta^{ijkl}=\eta^{jikl}=\eta^{ijlk}$,
since the metric is symmetric, $g_{ij}=g_{ji}$.
It can be written as a sum of the even and odd tensors $\eta_\erm$ and $\eta_\orm$ satisfying
\begin{align}
\eta_\erm^{ijkl}& = \eta_\erm^{klij},&
\eta_\orm^{ijkl}& = -\eta_\orm^{klij}.
\end{align}
The symmetric part includes shear and bulk viscosities, as well as, at finite $\qv$, other even tensors which can be constructed using additionally the momentum $q_i$. In this paper we restrict our attention to the antisymmetric part of the viscosity tensor, known as Hall or odd viscosity \cite{Avron1995, Avron1997}. Much like for the conductivity, the dissipationless\footnote{Dissipation can arise from 
 $\sigma_\Hrm$ and $\eta_\orm$, if these are not even functions of $\omega$. However, this does not occur for  systems at equilibrium, like the chiral superfluids and superconductors we consider. Indeed, the relations~\labelcref{eq:relWsig,eq:relWeta} together with the fact that the functional derivatives in \cref{eq:defVertices} can be taken in any order imply $\sigma^{ij}(\omega)=-\sigma^{ji}(-\omega)$ and $\eta^{ijkl}(\omega)=-\eta^{klij}(-\omega)$.} odd viscosity is a signature of the breaking of time-reversal symmetry. It has been shown in \cite{Golan2019} that, in an isotropic system that is symmetric under the combination of parity and time-reversal symmetries ($PT$ symmetry) the Hall viscosity tensor is fixed by only two independent components $\eta_\orm^{(1)}$ and $\eta_\orm^{(2)}$\footnote{Our definition of the viscosity tensor $\eta$ in \cref{eq:defViscLinResp} follows the standard hydrodynamic convention. It agrees with Refs.\cite{Bradlyn2012,Hoyos2013} and is opposite to that used in \cite{Golan2019}. Our definition \cref{Hvis} of the viscosity coefficients from $\eta$ follows \cite{Golan2019}. In particular, comparison with Refs.\cite{Bradlyn2012,Hoyos2013} is obtained via  $\eta^{H}=\eta_{H}=-\eta_{\text{o}}^{(1)}$.} 
\begin{equation} \label{Hvis}
\eta_\orm(\omega,\qv) =  \eta_\orm^{(1)}(\omega,\qv^2) \sigma^{xz} +  \eta_\orm^{(2)}(\omega,\qv^2)[(q_x^2-q_y^2)\sigma^{0x}-2q_x q_y \sigma^{0z}] 
\end{equation} 
where the $\sigma^{ab}$ matrices are antisymmetrized tensor products of the Pauli matrices $\sigma^a$
\begin{equation} \label{ansymP}
(\sigma^{ab})^{ijkl} = (\sigma^a)^{ij}(\sigma^b)^{kl} - (\sigma^b)^{ij}(\sigma^a)^{kl}.
\end{equation}

Since the momentum-dependent tensor that multiplies $\eta_{\orm}^{(2)}$ in \cref{Hvis} vanishes at $\qv=0$, one expects the response to the homogenous ($\qv=0$) perturbation to be fully determined by the first term proportional to $\eta_{\orm}^{(1)}$~\cite{Avron1997}, often identified in the literature with the Hall viscosity~\cite{Avron1995,Read2009,Bradlyn2012}. However, as we discuss later in \cref{sec:EFT,sec:microTheory}, in some cases the coefficient $\eta_{\orm}^{(2)}$ is singular like $1/\qv^2$ in the $\qv\to 0$ limit, such that $\eta^{ijkl}_\orm(\omega,\qv=0) \neq \lim_{\qv \to 0} \eta^{ijkl}_\orm(\omega,\qv)$. Indeed, when this happens the second term in \cref{Hvis} doesn't vanish at small but finite $\qv$ and cannot be dropped carelessly, while the viscosity in the homogenous limit is determined by the single coefficient $\eta_{\orm}^{(1)}$.

\subsection{Galilean Ward identities}
\label{sec:WardProper}

In a Galilean-invariant system composed of a single species  of particles, the conductivity and viscosity are not independent, as the transport of electrical charge is tied to the transport of momentum density. The formal expression of  this statement 
comes from the  Ward identities relating the correlation fuctions of the current $J^i$ and the stress tensor $T^{ij}$ to each other~\cite{taylor2010viscosity,Bradlyn2012,Geracie2015}.

Consider a non-relativistic theory whose action $S$ is invariant under global $\Uone$ transformations as well as under global spatial translations. By coupling  to an external $\Uone$ gauge field $A_\mu$ and defining the theory in  space with a spatial metric $g_{ij}$, we can promote these two global symmetries to local gauge invariance, provided $A_\mu$ and $g_{ij}$ transform as \cite{Son2006}
\begin{align}
\delta A_t &=	-\partial_t \alpha - \xi^k \partial_k A_t - A_k \partial_t \xi^k ,
\label{eq:transfLawAt}\\
\delta A_i &=  -\partial_i \alpha - \xi^k \partial_k A_i - A_k \partial_i \xi^k + m g_{ik}\partial_t \xi^k,
\label{eq:transfLawAi}\\
\delta g_{ij} &= -\xi^k\partial_k g_{ij}- g_{ik}\partial_j \xi^k-g_{kj}\partial_i\xi^k
\label{eq:transfLawg}
\end{align}
under infinitesimal $\Uone$ gauge transformations and spatial diffeomorphisms with respective parameters $\alpha(x)$ and $\xi^i(x)$.
 The induced action inherits the symmetries of the original action and as a result, 
\begin{equation}
\Wc[A_\mu+\delta A_\mu, g_{ij} + \delta g_{ij}] = \Wc[A_\mu, g_{ij}]. \label{eq:invarEffAct}
\end{equation}

As the invariance is valid for any infinitesimal transform, \cref{eq:invarEffAct} implies  two independent identities that holds at any point in spacetime and any value of the sources. Expressed in terms of expectation values, these read
\begin{align}
 \dfrac{1}{\sqrt{g}}\partial_t(\sqrt{g} \braket{J^t}) + \nabla_i \braket{J^i}&=0,\label{eq:continuityCurrent}\\
\dfrac{1}{\sqrt{g}} m \partial_\tau (\sqrt{g} \braket{J_k})  + \nabla_i \braket{T^i_k} 
&=E_k \braket{J^t} + \veps_{ik}\braket{J^i} B. \label{eq:continuityTens}
\end{align}
\cref{eq:continuityCurrent,eq:continuityTens} are respectively the continuity equations for the $\Urm(1)$ current and momentum density in the background of a general $\Urm(1)$ gauge field $A_\mu$ and the metric $g_{ij}$.  Here we introduced the covariant Levi-Civita derivative $\nabla_i$, the Levi-Civita tensor    $\veps_{ij}=\sqrt{g}\epsilon_{ij}$, $\veps^{ij}=(1/\sqrt{g})\epsilon^{ij}$, the electric field $E_j=\partial_t A_j-\partial_j A_t$  and the magnetic field $B = \veps^{ij}\partial_i A_j$.

Since \cref{eq:continuityCurrent,eq:continuityTens} are valid for any configurations of the sources, it is possible to take further derivatives to obtain relations between $n$-point correlation functions. For the two-point functions we derive these in \cref{app:ward}. As a result, we find a relation between the transport coefficients\footnote{As one can see in \cref{app:ward}, to get \labelcref{eq:WardMaster} we must replace the elastic tensor $\lambda^{ijkl}(\omega, \qv)$ with its $\qv=0$, $\omega=0$ expression; the Ward identity is thus valid for all momenta and frequencies up to  finite $\qv$ corrections that originate from $\lambda^{ijkl}(\omega, \qv)$.} 
\begin{equation}
m^2 (\omega^{+})^2 \sigma^{ij} (\omega,\qv) = q_k q_l \eta^{ikjl}(\omega,\qv) - \dfrac{1}{\ir \omega^{+}} q^i q^j \kappa^{-1}
\label{eq:WardMaster}
\end{equation}
where $\kappa^{-1}=-V \big(\partial P/ \partial V \big)_{S, N}$ is the inverse compressibility.  Projecting it on the antisymmetric part gives
\begin{equation} \label{WardH}
m^2 (\omega^{+})^2 \sigma_\Hrm (\omega,\qv^2) = -\qv^2 [\eta_{\orm}^{(1)}(\omega,\qv^2) -\qv^2 \eta_{\orm}^{(2)}(\omega,\qv^2)]
\end{equation}
which is valid for all $\qv$, $\omega$ provided $\lambda^{ijkl}(\omega, \qv)$ has no odd part.

\section{Effective field theory}
\label{sec:EFT}
At low energies and long wave-lengths the collective degrees of freedom of a nonrelativistic superfluid are determined by spontaneous symmetry breaking and the Galilean-invariant dynamics can be encoded in a non-linear effective action of Goldstone bosons \cite{Greiter:1989qb, Son2006}. The chiral ground state of two-dimensional fermions paired in the $l^{\text{th}}$ partial wave has the order parameter $\langle \psi_{\mathbf{q}} \psi_{-\mathbf{q}} \rangle\sim\Delta_{\mathbf{q}}=(q_x+i q_y)^l|\Delta_0|$ \cite{Read2000, volovikbook} which has a non-trivial phase winding around the Fermi surface.\footnote{Due to the Pauli principle, the chirality parameter $l$ for spinless fermions must be odd, while that of spin-full fermions where $\langle \psi_{\uparrow, \mathbf{q}} \psi_{\downarrow,-\mathbf{q}} \rangle\sim\Delta_{\mathbf{q}}=(q_x+i q_y)^l|\Delta_0|$ must be even.} As a result, the global particle number $\Urm(1)_N$ symmetry and the rotation $\SO(2)_R$ symmetry are both spontaneously broken by the condensate, while a special linear combination $\Urm(1)_D$ of these two symmetries leaves the order parameter invariant. The spontaneous symmetry breaking thus has the form $\Urm(1)_N\times \SO(2)_R \to \Urm(1)_D$ implying that there is only one Goldstone mode in the energy spectrum.   The effective field theory (EFT) of this Goldstone boson in a Galilean-invariant two-dimensional chiral superfluid was developed in \cite{Hoyos2013, Moroz2014a, Golan2019}. 

We first briefly review this theory in \cref{sec:reminderChiralSF} and present the Hall conductivity and Hall viscosity that were extracted from it. Subsequently, we develop in \cref{sec:EFTintrinsticSC} the effective theory of a two-dimensional intrinsic superconductor by incorporating the effects of the instantaneous Coulomb interaction and extract from the induced action the Hall responses of these chirally paired superconducting states. Finally,  in  \cref{sec:EFTmixedSC} we extend the calculation to the case of general long-ranged interactions, encompassing the case of mixed-dimensional chiral superconductors. A comprehensive analysis of linear response in chiral superconductors is presented in \cref{app:LR}.

\subsection{Chiral superfluid}
\label{sec:reminderChiralSF}
To first order in derivatives, the effective action of the Goldstone field $\theta$, coupled to a background spatial metric $g_{ij}$ and $\Urm(1)_{N}$ gauge field $A_\mu$, is fixed by the thermodynamic pressure as a function of chemical potential $P(\mu)$ by 
\beq \label{EFT:ac}
S[\theta;A,g]=\int_x \sqrt{g} P(X),
\eeq
where $X=D_{t} \theta-\frac{g^{i j}}{2m} D_{i} \theta D_{j} \theta$ and the covariant derivative
$D_{\mu} \theta = \partial_{\mu} \theta-A_{\mu}-s \omega_{\mu}$ \cite{Son2006, Hoyos2013}. Here the chirality parameter is $s=\pm l/2$ for a chiral superfluid paired in the $l^{\text{th}}$ partial wave, and the spin connection $\omega_\mu$ is constructed from a pair of orthonormal vielbein vectors\footnote{Out of the vielbein pair $e^a_i$ one can construct the spatial metric $g_{i j}=e_{i}^{a} e_{j}^{a}$ and the Levi-Civita tensor $\varepsilon_{i j}=\epsilon^{a b} e_{i}^{a} e_{j}^{b}$. Both of these tensors are invariant under local $\SO(2)_\nu$ rotation of vielbeins in internal space labeled by the index $a$. Hence, for a given metric $g_{ij}$ the vielbeins are not uniquely defined, and while we write for clarity the action \labelcref{EFT:ac} as a functional of the metric, it should rather read $S[\theta;A,e]$.} $e^a_i$ as
\begin{align} \omega_{t} & = \frac{1}{2}\left(\epsilon^{a b} e^{a j} \partial_{t} e_{j}^{b}+B\right), &
\omega_{i} & =  \frac{1}{2} \epsilon^{a b} e^{a j} \nabla_{i} e_{j}^{b}=\frac{1}{2}\left(\epsilon^{a b} e^{a j} \partial_{i} e_{j}^{b}-\varepsilon^{j k} \partial_{j} g_{i k}\right). \label{EFT:spincon}
\end{align}
The superfluid density is given by $n=-\left( 1/\sqrt{g}\right) \delta S / \delta A_{t}=P'(X)$, where the prime indicates a derivative. In the ground state  $\theta=\mu t + \text{const.}$, in which case $X$ reduces to the chemical potential $\mu$ and the density $n$ reduces to the thermodynamic expression $n_{0}=P'(\mu)$. 

Following a standard linear response calculation, the Hall transport coefficients were extracted from the EFT \labelcref{EFT:ac} in \cite{Hoyos2013, Moroz2014a, Golan2019}. The Hall conductivity was found to be equal to
\beq \label{condSF}
\sigma_{\Hrm}(\omega, \mathbf{q}^2)=\frac{s n_0}{2m^2} \frac{-\mathbf{q}^{2}}{\omega^{2}-c_{s}^{2} \mathbf{q}^{2}},
\eeq
where $c_s=\sqrt{\partial P/ \partial n}$ is the speed of sound. The two independent components of the Hall viscosity tensor \labelcref{Hvis} were calculated,\footnote{The relative sign between $\eta_{\text{o}}^{(1)}$ and $\eta_{\text{o}}^{(2)}$ in \cref{EFT:eta12cs} corrects a typo in Ref.\cite{Golan2019}.}
\beq \label{EFT:eta12cs}
\eta_{\mathrm{o}}^{(1)}(\omega, \mathbf{q}^2)=\frac{s n_{0}}{2}, \quad \eta_{\mathrm{o}}^{(2)}(\omega, \mathbf{q}^2)=-\frac{s n_{0}}{2} \frac{c_{s}^{2}}{\omega^{2}-c_{s}^{2} \mathbf{q}^{2}}.
\eeq

The action \labelcref{EFT:ac} is invariant under local $\Urm(1)$ gauge transformations and spatial diffeomorphisms \cite{Hoyos2013, Moroz2014a, Golan2019} resulting in the Ward identity \labelcref{WardH} that relates the Hall conductivity and viscosity tensors. In the homogeneous limit $\mathbf{q}\to0$ the Hall viscosity tensor $\eta_{\mathrm{o}}(\omega)$ \labelcref{Hvis} reduces to only one component $\eta_{\mathrm{o}}^{(1)}(\omega)$ and the odd version of the conductivity-viscosity Ward identity takes the simple form \labelcref{Wardsup}. 

It is clear from \cref{EFT:eta12cs} that the component $\eta_{\mathrm{o}}^{(2)}(\omega, \mathbf{q}^2)$ of the Hall viscosity tensor becomes formally singular at $\mathbf{q}=0$ in the incompressible limit $c_s\to \infty$. We will see in the following that a similar singularity arises in a chiral two-dimensional intrinsic superconductor.

\subsection{Intrinsic two-dimensional chiral  superconductor}
\label{sec:EFTintrinsticSC}
In order to incorporate an instantaneous logarithmic Coulomb interaction between fermions, we couple the superfluid effective theory to a mediating, or Hubbard-Stratonovich, scalar field $\chi$. The effective action of an intrinsic two-dimensional chiral superconductor is then 
\beq \label{EFT:acsc}
S[\theta,\chi;A,g]=\int_x \sqrt g \bigg[ P(X-\chi)+\bar n \chi+\frac {1} {2e^2} g^{ij}\partial_i \chi \partial_j \chi \bigg].
\eeq
The uniform  background density $\bar n$  ensures that the overall system is electrically neutral, and $e$ corresponds to the electric charge of a microscopic fermion. The equation of motion for $\chi$ is the two-dimensional Poisson equation \beq
\frac{1}{\sqrt{g}}\partial_{i}\left( g^{ij}\sqrt{g}\partial_{j} \chi \right)=-e\delta Q,
\eeq
where $  \delta Q= e[P'(X-\chi)-\bar n]$ is the total charge density. 

Since the Coulomb potential is instantaneous, it does not break Galilean symmetry. More generally, we impose that the Coulomb field $\chi$ transforms as a scalar under spatial diffeomorphisms and does not transform under local $\Urm(1)_{N}$ transformations. As a result, the action \labelcref{EFT:acsc} is invariant under both  transformations, and the Ward identity \labelcref{WardH} remains intact.

We now turn to the computation of linear response functions, namely the Hall conductivity and viscosity, based on the action \cref{EFT:acsc}.  We perform the computation within the random phase approximation (RPA), and to leading order in derivatives. These approximations amount to a quadratic expansion of  the action \cref{EFT:acsc} in all fields, and are discussed in  \cref{app:approximations}. We write the Goldstone field $\theta=\mu t- \varphi$, where the first term represents the ground state contribution, while the second term denotes the fluctuating part of the field. We define the Lagrangian density $\mathcal{L}$ by $S=\int_x \sqrt{g} \mathcal{L}$. Expanding $\mathcal{L}$  to quadratic order in the fluctuations $\varphi$ and the Coulomb field $\chi$, we find
\begin{align}
\mathcal{L}={}&{}P(\mu)-P'(\mu)\bigg(\chi+D_t \varphi+\frac{g^{ij}}{2m} D_i \varphi D_j \varphi\bigg) \nnl
{}&{}+\frac 1 2 P''(\mu)\big[\chi^2+(D_t \varphi)^2+2 \chi D_t \varphi\big]+\bar n \chi+\frac {1} {2e^2} g^{ij}\partial_i \chi \partial_j \chi,
\end{align}
where $D_\mu \varphi=\partial_\mu \varphi+A_\mu+s \omega_\mu$  and primes denote derivatives with respect to the chemical potential. Introducing $P_0=P(\mu)$, $n_0=P'(\mu)$ and $P''(\mu)=n_0/m c_s^2$, applying the charge neutrality condition $n_0=\bar n$ and rearranging the terms, we get
\beq \label{Lq}
\mathcal{L}=P_0-n_0 D_t \varphi-\frac{n_0}{2m} g^{ij} D_i \varphi D_j \varphi+\frac 1 2 \frac{n_0}{m c_s^2} (D_t \varphi)^2+ \frac{n_0}{m c_s^2} \chi D_t \varphi +\frac {1} {2e^2} g^{ij}\partial_i \chi \partial_j \chi+ \frac 1 2 \frac{n_0}{m c_s^2} \chi^2.
\eeq

At this stage we would like to compute the induced action $\Wc[A, g]$ by preforming the Gaussian functional integration over the Coulomb and Goldstone fields $\chi$ and $\varphi$.\footnote{ Gaussian functional integration in the presence of a background metric $g_{ij}$ is briefly summarized in \cref{app:gaus}.} First, we integrate over the Coulomb field $\chi$ and find

\begin{align}
\mathcal{L}={}&{} P_0 -n_0 D_t \varphi-\frac{n_0}{2m} g^{ij} D_i \varphi D_j \varphi\nnl
{}&{}+\frac 1 2 \frac{n_0}{m c_s^2} (D_t \varphi)^2 +\frac{1}{2}\bigg(\frac{n_0}{m c_s^2} D_t \varphi \bigg) \frac{1}{\boldsymbol\nabla^2/e^2-n_0/m c_s^2} \bigg(\frac{n_0}{m c_s^2} D_t \varphi \bigg) \nnl
={}&{}P_0-n_0 D_t \varphi-\frac{n_0}{2m} g^{ij} D_i \varphi D_j \varphi+\frac 1 2 \frac{n_0}{m} D_t \varphi
\tilde c_s^{-2} 
D_t \varphi, \label{eq:lagPhiEftCoul}
\end{align}
where we introduced the renormalized momentum-dependent speed of sound operator as 
\begin{equation}
    \cst^{2}=\cs^{2}- \omega_p^2/\boldsymbol\nabla^2,
\end{equation}
with $\omega_p=\sqrt{e^2 n_0/m}$ the plasma frequency. The Lagrangian \labelcref{eq:lagPhiEftCoul} can alternatively be obtained following the derivation in \cref{app:legendre}.
 We are now ready to integrate out the Goldstone field $\varphi$. First, we put the functional integral into the standard Gaussian form by performing several integrations by parts. Next, we follow \cref{app:gaus} and obtain the induced action
\begin{align}
\Wc[A,g]={}&\int_x \sqrt{g} \bigg\{P_0 -n_0\left(\mathsf{A}_{t}+\frac{g^{i j}}{2 m} \mathsf{A}_{i} \mathsf{A}_{j}\right)+\frac{1}{2} \frac{n_0}{m} \mathsf{A}_{t} \tilde c_{s}^{-2} \mathsf{A}_{t} \nnl
{}&{}-\frac{n_0}{2m}\left[m \dot{f}+\boldsymbol\nabla \cdot \boldsymbol{\mathsf{A}}-\tilde \partial_t \tilde c_{s}^{-2} \mathsf{A}_{t}\right] \frac{1}{\boldsymbol\nabla^{2}- \tilde \partial_t \tilde c_{s}^{-2} \partial_{t}}\left[m \dot{f}+\boldsymbol\nabla \cdot \boldsymbol{\mathsf{A}}-\tilde \partial_t \tilde c_{s}^{-2} \mathsf{A}_{t}\right]
 \bigg\},\label{eq:indac2dsc}
\end{align}
where we introduced $f=\log \sqrt{g}$, $\tilde \partial_t=\partial_t+\dot f$ and $\mathsf{A}_\mu=A_\mu+s \omega_\mu$.  Expanding now the induced action around flat space and following  \cref{subsec:Induced-action-and}, the induced action can be written in the covariant form in Fourier space
\begin{align}
\Wc[A,g]= & \int_x \Big[2 P_{0} h-n_{0}\mathsf{A}_{t} \nnl {}&{}+\dfrac{1}{2}\dfrac{n_{0}}{m}\dfrac{\cst^{2}\Bsf^{2}-\Esfv^{2}+(\ir s/m)\mathsf{E}^{i}q_{i}B-(s^{2}/4m^{2})\qv^{2}B^{2}}{\omega^{2}-\cst\qv^{2}}\nonumber \\
 & +2n_{0}\dfrac{\tilde{c}_{s}^{2}h[\ir q_{i}\mathsf{E}^{i}+(s/2m)\qv^{2}B]-m\cst^{2}\omega^{2}h^{2}}{\omega^{2}-\cst^{2}\qv^{2}}\Big], \label{eq:indActEFTomri}
\end{align}
where $h_{ij}=g_{ij}-\delta_{ij}$, $h= h_i^i$, and $\mathsf{E}_i = \partial_t \mathsf{A}_i - \partial_i \mathsf{A}_t$ and $\mathsf{B}= \varepsilon^{ij}\partial_{i}\mathsf{A}_j$ are the electric and magnetic fields constructed with $\mathsf{A}_{\mu}$. As we argue in \cref{app:LR}, the induced action of the chiral superconductor is identical to the one found for the chiral superfluid provided the renormalized speed of sound $\tilde c_s$ is used. 

It is straightforward now to extract electromagnetic and gravitational linear responses  from either \labelcref{eq:indac2dsc} or \labelcref{eq:indActEFTomri}. Their comprehensive calculation and analysis  based on the latter expression of the induced action is performed in \cref{subsec:Induced-action-and}.  Here we present results for the Hall conductivity and viscosities. Using \cref{eq:relWsig} we find the Hall conductivity in flat space and $A_\mu=0$
\beq \label{conSC}
\sigma_{\Hrm}(\omega, \mathbf{q}^2)=\frac 1 2 \epsilon_{ij} \sigma^{ij}=\frac{s n_0}{2m^2} \frac{-\mathbf{q}^{2}}{\omega^{2}-\omega_p^2-c_{s}^{2} \mathbf{q}^{2}}.
\eeq
This result resembles the Hall conductivity of the chiral superfluid \labelcref{condSF} with the only difference in the denominator stemming from the gapped nature of the  collective plasmon mode. As a result, at small frequency $\omega$ and momentum $\mathbf{q}$ the Hall conductivity vanishes as a quadratic function of the momentum
\begin{equation}
\sigma_\Hrm(\omega\to 0,\qv\to 0) =  \dfrac{s}{2 m e^2} \qv^2.
\end{equation}
In contrast to the Hall conductivity of the gapless superfluid \labelcref{condSF}, this result is unique and does not depend on the order of limits.

Using now \cref{eq:relWeta,Hvis} we extract the two independent components of the Hall viscosity tensor
\beq \label{EFT:eta12csSC}
\eta_{\mathrm{o}}^{(1)}(\omega, \mathbf{q}^2)=\frac{s n_{0}}{2}, \quad \eta_{\mathrm{o}}^{(2)}(\omega, \mathbf{q}^2)=-\frac{s n_{0}}{2} \frac{1}{\mathbf{q}^2} \frac{\omega_p^2+c_{s}^{2} \mathbf{q}^2}{\omega^{2}-\omega_p^2-c_{s}^{2} \mathbf{q}^{2}}.
\eeq
We find that the component $\eta_{\mathrm{o}}^{(1)}$ is identical to the one found for the chiral superfluid \labelcref{EFT:eta12cs}. In other words, the instantaneous Coulomb interaction does not affect $\eta_{\mathrm{o}}^{(1)}$. On the other hand, the component $\eta_{\mathrm{o}}^{(2)}$ is modified. Most notably, in the homogeneous limit $\mathbf{q}\to 0$ it diverges as $1/\mathbf{q}^{2}$. We thus conclude that due to the long-range Coulomb potential the homogeneous limit of the Hall viscosity tensor \labelcref{Hvis} is ill-defined since the result depends on the direction of the vanishing vector $\mathbf{q}$. We attribute this peculiar singularity to the instantaneous nature of the Coulomb potential, and in \cref{sec:conc} will provide an outlook on its fate in a model which involves photons that propagate with a finite speed of light.

It is straightforward to check that the Hall conductivity and viscosity found above satisfy the Ward identity \labelcref{WardH} that follows from the Galilean symmetry of the chiral superconductor, where the electromagnetic interaction is approximated by the instantaneous Coulomb interaction. In the regime of small momentum $\mathbf{q}$, in contrast to the chiral superfluid, the contribution of $\eta_{\mathrm{o}}^{(2)}$ to the Ward identity does not drop.
The Hall conductivity encodes information about a particular combination of the two independent components of the Hall viscosity tensor, but is not sufficient to fix either of them separately.

\subsection{Mixed-dimensional chiral  superconductor}
\label{sec:EFTmixedSC}
In a mixed-dimensional superconductor the Coulomb potential decays as $1/|\rv|$ at large distances which is dictated by the three-dimensional nature of the electromagnetic field. It is straightforward to generalize the effective theory developed in the previous section to this case. To include a generic power-law decaying interaction,  we start from the effective action
\beq \label{EFT:acscm}
S[\theta,\chi;A,g]=\int_x \sqrt g \bigg[ P(X-\chi)+\bar n \chi+\frac {1} {2e^2} \chi (-\nablav^2)^{\alpha/2} \chi \bigg],
\eeq
where $\nablav^2$ is the covariant Laplace operator and $0\le \alpha \le 2$. In flat space the scalar field $\chi$ mediates an instantaneous repulsive central  potential that decays as $|\rv|^{\alpha-2}$. The special case $\alpha=1$ corresponds to the mixed-dimensional superconductor. The case $\alpha=2$ is the intrinsic superconductor discussed in the previous subsection, where the Coulomb potential is logarithmic.

Following the steps of the previous subsection one arrives at the induced action \labelcref{eq:indac2dsc}, where now the inverse square of the renormalized speed of sound is
\beq
\tilde c_s^{-2}=\frac{(-\nablav^2)^{\alpha/2}}{c_s^2(-\nablav^2)^{\alpha/2}+ \omega_p^2}.
\eeq
Here as before we introduced $\omega^2_p=e^2 n_0/m$. We stress that $\omega_p$ is a frequency that defines the plasmon gap only in the case of the intrinsic superconductor, i.e., for  $\alpha=2$. For all $0<\alpha<2$ plasmon collective modes are gapless and the units of $\omega_p$ are $[\omega q^{(\alpha-2)/2}]$.

The induced action defined by \cref{eq:indac2dsc} contains all necessary information to extract electromagnetic and gravitational linear responses. This is discussed in detail for a generic value of $\alpha$ in \cref{subsec:Induced-action-and}. Here we present the Hall responses. The Hall conductivity
\beq \label{conSCgen}
\sigma_{\Hrm}(\omega, \mathbf{q}^2)=\frac 1 2 \epsilon_{ij} \sigma^{ij}=\frac{s n_0}{2m^2} \frac{-\mathbf{q}^{2}}{\omega^{2}- \omega_p^2|\mathbf{q}|^{2-\alpha} -c_{s}^{2} \mathbf{q}^{2}}.
\eeq
In a mixed-dimensional superconductor ($\alpha=1$) the plasmon mode is gapless and disperses as $\sqrt{|\mathbf{q}|}$ as low momenta. This implies that the limits $\omega\to 0$ and $\mathbf{q}\to 0$ of the Hall conductivity do not commute. The Hall viscosities extracted from the induced action are
\beq \label{EFT:eta12csSCmixed}
\eta_{\mathrm{o}}^{(1)}(\omega, \mathbf{q}^2)=\frac{s n_{0}}{2}, \quad \eta_{\mathrm{o}}^{(2)}(\omega, \mathbf{q}^2)=-\frac{s n_{0}}{2} \frac{1}{|\mathbf{q}|^\alpha} \frac{\omega_p^2+c_{s}^{2} |\mathbf{q}|^\alpha}{\omega^{2}-\omega_p^2|\mathbf{q}|^{2-\alpha}-c_{s}^{2} \mathbf{q}^{2}}.
\eeq
The component $\eta_{\mathrm{o}}^{(2)}$ diverges as $|\mathbf{q}|^{-\alpha}$ in the homogeneous $\mathbf{q}\to 0$ limit. We thus conclude that for any $0\le\alpha<2$ the $\eta_{\mathrm{o}}^{(2)}$ contribution to the Hall viscosity \labelcref{Hvis} tensor vanishes for $\mathbf{q}\to 0$.
As a result, the gapped intrinsic superconductor, where the Hall viscosity tensor is ill-defined in the limit $\mathbf{q}\to 0$, is an exceptional case.

An explicit calculation confirms that the conductivity-viscosity Ward identity \labelcref{WardH} is satisfied for a generic value of $\alpha$. In the homogeneous case, if $\alpha\ne 2$, this Ward identity simplifies to the form \labelcref{Wardsup}.

\section{Microscopic theory}
\label{sec:microTheory}

In this Section, starting from a microscopic theory of fermions we derive the transport coefficients of the chiral paired states. First, we consider the case of the neutral chiral superfluid, where spinless fermions attract each other via a short-range potential. Next, we study the superconductor by including into the microscopic model the effects of the long-range repulsive Coulomb interaction.

The starting point of our discussion is the action of spinless fermions $\psi$ interacting via a separable \textit{p}-wave short-range potential in space with arbitrary metric $g_{ij}$ and in presence of a $\Uone_N$ gauge field $A_\mu$,
\begin{equation}
S = \int_{x} \sqrt{g} \bigg\{\psi^*\bigg[\partial_\tau - \ir  A_0 + \dfrac{g^{ij}(p_i- A_i)(p_j- A_j)}{2m}-\mu\bigg]\psi
- \lambda g^{ij} (\psi^* p_i \psi^*)(  \psi p_j \psi) \bigg\}, \label{eq:microAct1}
\end{equation}
where $x=(\tau,\rv)$ is the $(2+1)$-dimensional space-time coordinate,   $p_i=-\ir\nabla_i$  the momentum operator, $m$ the fermion mass 
and $\lambda$ the interaction strength. Note that the $\Uone_N$ gauge field doesn't appear in the interaction term due to the Pauli principle. 

In \cref{sec:sub_BCS}, we derive from \cref{eq:microAct1} the  Bardeen–Cooper–Schrieffer (BCS) theory of \textit{p}-wave pairing. An induced action for the  external fields is formally obtained in \cref{sec:sub_ind}, from which we compute the transport coefficients in \cref{sec:sub_transNoCoul}. In \cref{sec:sub_Coul}, we extend these results in presence of long-range interactions such as the Coulomb interaction.
  
In this Section the microscopic action \cref{eq:microAct1} is formulated in Euclidian (imaginary) time, $t\to-\ir \tau$, which is more convenient for the linear response computation.
The induced action is given by the functional integral
\begin{equation}
\exp(-\Wc[A,g]) = \int \Dc[\psi,\psi^*]\exp(-S[\psi,\psi^*;A,g])
\end{equation}
and like its real-time counterpart, it is the generating functional of the connected correlation functions. We compute two-point correlation functions $f(\ir \omega_n,q_x,q_y)$ depending on a Matsubara frequency $\ir \omega_n$ from which the real-time, retarded dynamical correlation functions $f^\Rrm(\omega,\qv)$ are obtained after analytic continuation $\ir \omega_n \to \omega^+ = \omega + \ir 0^+$. Notice that upon going to imaginary time, the time component of the $\Uone_N$ gauge field $A_t$ transforms like a time derivative, $A_t \to \ir A_0$. This should be taken into account when going back to real time to determine the retarded correlation functions.

\subsection{BCS action}
\label{sec:sub_BCS}

To decouple the $p$-wave interaction, we introduce an auxiliary two-component bosonic (complex) field $\overline{\Deltav}$, and perform a Hubbard-Stratanovitch transform to get
\begin{align}
S[\psi,\psi^*,\overline{\Deltav};A,g] ={}& \int_x\sqrt{g}\bigg\{ \psi^* \bigg[\partial_\tau-\ir  A_0 + \dfrac{g^{ij}(p_i-A_i)(p_j-A_j)}{2m}-\mu\bigg]\psi \nnl
{}&- \dfrac{1}{2}(\overline{\Delta}^i)^*(\psi p_i \psi) -\dfrac{1}{2}\overline{\Delta}^{i}(\psi^* p_i \psi^*)+\dfrac{1}{4\lambda}|\overline{\Deltav}|^2\bigg\}.
\label{eq:act_post_HS_interm}
\end{align}

In absence of external sources ($g_{ij}=\delta_{ij}$, $A_\mu=0$), the saddle point for the pairing field $\overline{\Deltav}$ is given by
\begin{align}
\overline{\Deltav} & = \Delta \er^{\ir \theta} \uv,&
\uv &= (1,\pm \ir) 
\end{align}
with $\Delta \geq 0$. $\Delta$ is the magnitude of the order parameter for the symmetry breaking pattern discussed in \cref{sec:EFT}, for the specific case of $p$-wave ($l=1$) superfluidity. While in the normal phase $\Delta=0$, in the superfluid regime $\Delta\neq 0$. We also define the global phase $\theta$ of the order parameter. The sign $\pm$ in $\uv$ distinguishes the two ground states with opposite chiralities, characterized by the angular momentum per particle $s=\pm 1/2$.
 Both the $\Uone_N$ gauge symmetry and the $\SO(2)_R$ rotation symmetry are spontaneously broken, while $\pv \cdot \overline{\Deltav}  $ remains invariant under the so-called diagonal $\Uone_\Drm$ symmetry. Furthermore, the ground states break down the time reversal ($\mathit{T}$) and parity ($\mathit{P}$) symmetries, while remaining invariant under the $\mathit{PT}$ combination.

Fluctuations of the pairing field around the saddle-point value correspond  to the four collective modes of the superfluid. In particular, the gapless phase mode is crucial to preserve the $\Uone_N$ gauge invariance of the theory. The other three modes are gapped and to investigate the theory at the BCS level, we discard their fluctuations while retaining the phase mode to Gaussian (quadratic) order.

For non-vanishing sources, the saddle point value of $\overline{\Deltav}$ depends on both $A_{\mu}$ and $g_{ij}$. We will now transform the action \labelcref{eq:act_post_HS_interm} in a manner where the $\Uone_N$ gauge invariance and diffeomorphism invariance are manifest at the mean-field level. First, since under $\Uone_N$ gauge transforms $\psi \to \exp(\ir \alpha)\psi$ the pairing field has to transform as $\overline{\Deltav} \to \exp(2\ir \alpha) \overline{\Deltav}$, we now decompose
\begin{equation}
\overline{\Deltav} = \Deltav \er^{2\ir \theta}
\end{equation}
with $\theta$ the fluctuating global phase of the pairing field, and perform the unitary transform $\psi \to \exp(\ir \theta)\psi$. That is equivalent to making the substitution in the action
\begin{align}
\overline{\Deltav}\to {} & \Deltav &
A_\mu \to {}& \Ac_\mu = A_\mu-\partial_\mu \theta
\end{align}
with each term being now manifestly $\Uone$ gauge invariant.

Furthermore, we introduce the a pair of vielbein vectors $e^{ia}$ which satisfy $e^{ia}\delta_{ab}e^{jb}=g^{ij}$. We now decompose $\overline{\Delta}^i = \overline{\Delta}_a e^{ai}$. In curved space, the mean-field configuration is given by $\overline{\Delta}_a\propto u_a=(1,\pm \ir)$~\cite{Read2000,PhysRevB.93.024521, PhysRevB.94.125137, Golan2019}.
 
 After these manipulations, the action reads
\begin{align}
S[\psi,\psi^*;\Ac,g] ={}& \int_x\sqrt{g}\bigg\{ \psi^* \bigg[\partial_\tau-\ir  \Ac_0 + \dfrac{g^{ij}(p_i-\Ac_i)(p_j-\Ac_j)}{2m}-\mu\bigg]\psi \nnl
{}&- \dfrac{1}{2}\Delta_a^*e^{ia}(\psi p_i \psi) -\dfrac{1}{2}\Delta_ae^{ia}(\psi^* p_i \psi^*)\bigg\}
\end{align} 
and the BCS approximation is obtained by setting $\Delta$ to its mean-field value, $\Delta_a = \Delta u_a$ with $\Delta>0$. We have discarded the constant $|\Deltav|^2/4\lambda$ contribution to the action which is important to determine the saddle point but doesn't further contribute to the calculation. As the vielbeins transform like vectors under general coordinate transformations, the mean-field action remains diffeomorphism-invariant. Combined with the $\Uone_N$ gauge invariance, this justifies the form of the mean-field action and implies that transport coefficients derived from this theory must satisfy the Ward identities \labelcref{eq:WardMaster,WardH}.

\subsection{Induced action}
\label{sec:sub_ind}

The action is quadratic in the fermion fields. We introduce the Nambu spinors
\begin{align}
\Psiv^\dagger_x&=(\psi^*_x,\psi_x), &
\Psiv_x&=(\psi_x,\psi^*_x)^{\mathsf{T}}
\end{align}
to rewrite the action
\begin{equation}
S[\Psiv,\Psiv^\dagger;\Ac,h] = -\dfrac{1}{2} \int_{x,x'}  \Psiv^\dagger_x \Gc_{x,x'}^{-1} \Psiv_{x'} \label{eq:effactForm}
\end{equation}
where $\Gc^{-1}$ is the inverse Nambu propagator, depending on $\Ac_\mu$ and $h_{ij}=g_{ij}-\delta_{ij}$. It reads
\begin{align}
\Gc^{-1}_{x,x'}={}&
\sqrt{g}\bigg\{-\partial_\tau \sigma^0
+ \ir \Ac_0 \sigma^z + \dfrac{1}{2}(\partial_\tau f) \sigma^z
 - \xi_{\pv,c}(\Ac,h)
+\Delta  e^{i}_a p_i \sigt^a
\bigg\}
\delta(x-x').\label{eq:FullNambuProp}
\end{align}
In \cref{eq:FullNambuProp}, we introduce the shorthand notations $f=\log\sqrt{g}$,
\begin{equation}
\xi_{\pv,c}(A,h) =
\dfrac{g^{ij}}{2m}[p_ip_j \sigma^z-(p_i A_j+A_i p_j) \sigma^0 + A_iA_j \sigma^z] - \mu  \sigma^0 ,
\end{equation}
$\xi_{\pv,c}(A=0,h=0)=\xi_{\pv}\sigma^z$, $\xi_{\pv}=\pv^2/2m-\mu$  and the ``twisted'' Pauli matrices $\sigt$ defined by $\sigt_y = \mp \sigma_y$, $\sigt_a=\sigma_a$ for $a\neq y$, where $\mp$ is fixed by the chirality of the $p\pm \ir p$ state.

The inverse bare propagator $\Gc_0^{-1}$, obtained by dropping $h$ and $\Ac$, is diagonal in Fourier space,
\begin{align}
\mathcal{G}_{0,q}^{-1}&=
 \ir \omega_n \sigma_0 - \xi_\qv \sigma_z  + \Delta q_a \sigt^a
=
\begin{pmatrix}
\ir \omega_n  - \xi_\qv	&	\Delta \uv \cdot {\qv}\\
\Delta \uv^*\cdot {\qv}			& 	\ir \omega_n + \xi_\qv
\end{pmatrix},
\label{eq:MFpropag}
\end{align}
with $q=(\ir \omega_n,\qv)$. Inverting $\Gc_0^{-1}$ yields
\begin{equation}
\Gc_{0,q}=
\begin{pmatrix}
G_q	&	F_q\\
F_q^*			& 	- G_{-q}	
\end{pmatrix},
\end{equation}
where we introduce the normal and anomalous Green functions, $G_q$ and $F_q$ respectively,
\begin{align}
G_q &= -\braket{\psi_q\psi_q^*}=-\dfrac{\ir \omega_n+\xi_\qv}{ \omega_n^2+\Delta^2 {\qv^2}+\xi^2_\qv},&
F_q &= -\braket{\psi_q\psi_{-q}}=\dfrac{\Delta  \qv \cdot \uv }{ \omega_n^2+\Delta^2 {\qv^2}+\xi^2_\qv}.
\label{eq:defGF}
\end{align}
For future convenience, we also introduce
\begin{equation}
f_q = \dfrac{\Delta }{ \omega_n^2+\Delta^2 {\qv^2}+\xi^2_\qv}
\label{eq:defsmallf}
\end{equation}
such that $F_q=(\qv \cdot \uv) f_q$.

The next step is to integrate the fermions to obtain an induced action $S[\Ac,h]$ for the phase mode $\theta$ and the sources $A$ and $h$. Using the standard perturbation theory~\cite{Lutchyn2008}, we expand the inverse propagator in powers of the sources, writing $\Gc^{-1}=\Gc_0^{-1}-\Gamma$, with
\begin{equation}
\Gamma_{x,x'} = 
(1-\sqrt{g_x})\Gc^{-1}_{0,x,x'}-
\bigg\{
\Delta  (e^{i}_a-\delta^{i}_a) p_i \sigt^a  +\bigg[\ir  \Ac_0 + \dfrac{1}{2}\partial_\tau f\bigg] \sigma^z
 - (\xi_{\pv,c}(A,h)-\xi_{\pv}\sigma^z)
\bigg\}
\delta(x-x') \label{eq:defGammaGeneral}
\end{equation}
being a vertex representing the coupling of the fermions to the phase mode and the sources.

The effective action then reads up to second order in $h$ and $\Ac$
\begin{align}
S[\Ac,h]=-\Tr \ln (-\Gc_0^{-1} + \Gamma) = -\Tr \ln(- \Gc_0^{-1}) + \Tr \Gc_0 \Gamma +  \dfrac{1}{2} \Tr \Gc_0 \Gamma\Gc_0 \Gamma +o(\Ac^2,h^2)
\label{eq:loopExpansionSeff}
\end{align}
where the trace $\Tr$ runs over both the spinor indices and space-time coordinates.
Due to the presence of both sources $h$ and $A$, keeping track of all terms in the expansion is cumbersome. Because of this we first reorganize \cref{eq:loopExpansionSeff} in powers of $\Ac$,
\begin{equation}
S[\Ac,h]=S_0[h] + \int_x \Ac_{\mu,x} N^{\mu}_x[h] + \dfrac{1}{2} \int_{x,x'} \Ac_{\mu,x} Q^{\mu\nu}_{x,x'}[h]\Ac_{\nu,x'}
\label{eq:formalExpansionSeff}
\end{equation}
where $S_0[h]$, $N^{\mu}_x[h]$ and $Q^{\mu\nu}_{x,x'}[h]$ are functionals of the metric $h$ which can be inferred by identifying \cref{eq:loopExpansionSeff,eq:formalExpansionSeff}.
 By construction $Q^{\mu\nu}_{x,x'}[h]$ is symmetric, $Q^{\mu\nu}_{x,x'}[h]=Q^{\nu\mu}_{x'x}[h]$. 
 
 Now we integrate out the phase field $\theta$ to get the induced action
\begin{align}
\Wc[A,h]={}&\Wc_0[h] + \int_x A_{\mu,x} N^{\mu}_x[h] + \dfrac{1}{2} \int_{x,x'} A_{\mu,x} Q^{\mu\nu}_{x,x'}[h]A_{\nu,x'}\nnl
{}&-\dfrac{1}{2}\int_{x,x'} \partial_{x_\mu}\bigg( N^{\mu}_x[h] + \int_{y} Q^{\mu\nu}_{x,y}[h] A_{\nu,y}\bigg) [\partial_\mu \partial_\nu Q^{\mu\nu}]^{-1}_{x,x'}\nnl
{}&\times
\partial_{x'_\mu}\bigg( N^{\mu}_{x'}[h] + \int_{y} Q^{\mu\nu}_{x',y}[h] A_{\nu,y}\bigg), \label{eq:inducedActionFormal}
\end{align}
where $[\partial_\mu \partial_\nu Q^{\mu\nu}]^{-1}_{x,x'}$ is understood as the inverse of $\partial_{y_\mu}\partial_{y'_\nu}Q^{\mu\nu}_{y,y'}$ in the operator sense.

\subsection{Hall transport coefficients of a chiral superfluid}
\label{sec:sub_transNoCoul}

From the induced action \labelcref{eq:inducedActionFormal}, the transport coefficients $\sigma^{ij}$, $\eta^{ijkl}$ are deduced by computing functional derivatives of $\Wc$ wrt $A$ and $h$, see \cref{eq:relWsig,eq:relWeta}.

\subsubsection{Conductivity}
\label{sec:sub_condNoCoul}

\begin{figure}
    \centering 
    \begin{subfigure}[b]{0.41\textwidth}
    \centering 
    \includegraphics{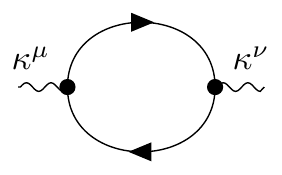}
        \caption{Bubble contribution to $Q^{\mu\nu}(q)$.}
        \label{fig:QmunuDiag}
    \end{subfigure}
    ~
    \begin{subfigure}[b]{0.41\textwidth}
    \centering
    \includegraphics{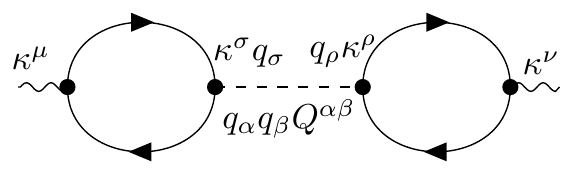}
        \caption{Phase mode contribution to $K^{\mu\nu}(q)$.}
        \label{fig:KmunuDiag}
    \end{subfigure}
\par

    \caption{Diagrammatic representations of the loop integral contributions to the correlators (a) $Q^{\mu\nu}(q)$ and (b) $K^{\mu\nu}(q)$. The solid line stands for the bare fermion Nambu propagator $\Gc_0$, the dashed line for the phase mode $\theta$ propagator, the wavy line for the external $\Uone_N$ gauge field $A_\mu$ insertions, and the dots for the interaction vertices $\kappa^\mu$, see \cref{eq:defQbubble}.}
\end{figure}

To determine the conductivity $\sigma^{ij}$, it is sufficient to investigate the theory in flat space $h=0$. In that case the calculation reduces to that in Ref.~\cite{Lutchyn2008}; we present its outline in \cref{app:MicCalcFSnoC}. To present the results in a compact way, and anticipating the analytic continuation, we make the replacement $\ir A_\tau \to A_\tau$ and introduce $q_0=-\ir \omega_n$ such that $\ir \Ac_{0,q} = \ir( A_{0,q} + \ir \omega_n \theta_q )$ becomes $\Ac_{0,q} = A_{0,q} - \ir q_0 \theta_q$. After that replacement,  the induced action for $\Ac$ becomes
\begin{equation}
S[\Ac] = 
\dfrac{1}{2}\int_q \Ac_{-q,\mu} Q^{\mu\nu}(q) \Ac_{q,\nu} 
\label{eq:action_gauge_phase_simpl}
\end{equation}
 with the correlation (polarization) functions $Q^{\mu\nu}(q)$ defined by
\begin{align}
Q^{\mu\nu}(q) &= \dfrac{n_0}{m}\delta^{\mu=\nu\neq 0} + \dfrac{1}{2} \int_p    \tr [\kappa^\mu_{p,q} \Gc_{0,p}\kappa^\nu_{p,q} \Gc_{0,p+q}],
&
\kappa^0_{p,q}&= \sigma^z,&
\kappa^j_{p,q}&=\dfrac{q^j+2p^j}{2m} \sigma^0,
\label{eq:defQbubble}
\end{align}
with $\tr$ denoting the trace only over the internal spinor indices and $\int_p=(2\pi)^{-3}\int \dr^2 \qv\,\dr \omega_n$ the summation over both momenta $\pv$ and Matsubara frequencies $\ir \omega_n$. While the first term on the rhs of \cref{eq:defQbubble} is the diamagnetic contribution to the current-current correlation function, the second term is a polarization bubble represented diagramatically in \cref{fig:QmunuDiag}.

Upon integrating out the phase mode $\theta$ we obtain the induced action for the $\Uone_N$ gauge field $A_\mu$,
\begin{align}
\Wc[A] &= 
\dfrac{1}{2}\int_q A_{-q,\mu} K^{\mu\nu}(q) A_{q,\nu},
\label{eq:action_gauge_phase_K} \\
K^{\mu\nu}(q) &= Q^{\mu\nu}(q) -  \dfrac{Q^{\mu\rho}(q)q_\rho q_\sigma Q^{\sigma\nu}(q)}{q_\alpha q_\beta Q^{\alpha\beta}(q)},
\label{eq:defK}
\end{align}
where the phase mode contribution to $K^{\mu\nu}(q)$ is represented in \cref{fig:KmunuDiag}. Since an arbitrary gauge transform reads $A_\mu \to A_\mu -\ir q_\mu \beta(q)$ for some scalar function $\beta$, $q_\mu K^{\mu\nu}(q)=K^{\mu\nu}(q)q_\nu=0$ enforces the $\Uone_N$ gauge invariance of the induced action. \cref{eq:action_gauge_phase_K} gives
\begin{align}
\Wc^{(2,0)ij}(q)=K^{ij}(q)
\end{align}
from which the conductivity $\sigma^{ij}(\omega,\qv)$, defined by \cref{eq:relWsig}, is deduced after analytic continuation,
\begin{equation}
\sigma^{ij}(\omega,\qv) = \dfrac{1}{\ir \omega^{+}} K^{\Rrm,ij}(\omega,\qv)
\end{equation}
where $K^{\Rrm,ij}(\omega,\qv)=K^{ij}(\ir \omega_n\to \omega^+,\qv)$ denotes the retarded part of the correlation function $K^{ij}(\ir \omega_n,\qv)$. The evaluation of the one-loops diagrams contributing to it is done in \cref{app:CalcQintegrals}. The resulting Hall conductivity, in the small momentum and frequency regime, is
\begin{equation}
\sigma_\Hrm (\omega,\qv^2)=\dfrac{s n_0}{2 m^2} \dfrac{-\qv^2}{\omega^2 - \cs^2 \qv^2}
\label{eq:exprSigHMicroNoC}
\end{equation}
where $s=\pm 1/2$ is the angular momentum per particle in the p-wave chiral ground state and $\cs = \sqrt{2\pi n_0/m^2}$ is the speed of sound. This result is in agreement with the effective field theory result  \labelcref{condSF} and the previous microscopic calculation~\cite{Lutchyn2008}.

In the above calculation, two ingredients are necessary to obtain a non-vanishing Hall conductivity.
First, preserving the $\Uone_N$ gauge invariance of the theory is crucial. Indeed, since $Q^{ij}(q)=Q^{ji}(q)$, $\sigma_\Hrm$ would vanish if we hadn't kept the phase mode $\theta$. Furthermore, we stress the role of the current-density correlation function, $Q^{0i}(q)$, which is a sum of an even and odd parts, see \cref{app:eqDefQevenodd},
\begin{align}
Q^{0i}(q) &= Q^{0i}_\er(q) + Q^{0i}_\orm(q),& 
Q^{0i}_\er(q) &= Q^{0i}_\er(-q),&
Q^{0i}_\orm(q) &= -Q^{0i}_\orm(-q).
\end{align}
In particular, while the density-density $Q^{00}$, current-current $Q^{ij}$ and even current-density $Q^{0i}_\er$ correlators are defined in a similar manner as in a non-chiral superfluid (up to the precise form of the gap function), the presence of the odd current-density correlator $Q^{0i}_\orm$ is only possible due to the time-reversal symmetry breaking. 
For $\sigma_\Hrm(q)$ to be finite, it is necessary to have a nonvanishing $Q^{0i}_\orm(q)$. Changing the chirality of the ground state from $\qv\cdot\uv=q_x \pm \ir q_y$ into $q_x\mp \ir q_y$ flips the sign of $Q^{0i}_\orm(q)$ implying $\sigma_\Hrm\to -\sigma_\Hrm$.

\subsubsection{Viscosity}
 \label{sec:sub_viscNoCoul}

To determine the Hall viscosity, it is sufficient to work with the induced action \labelcref{eq:inducedActionFormal} with $h_{ij}\neq 0$ and $A_\mu=0$. Only the fist and last term of \cref{eq:inducedActionFormal} remain. Furthermore, as a consequence of \cref{eq:flatSpaceTrivialLoops} $\partial_{x^{\mu}}N_x^\mu[h]= \Oc(h)$, so it is enough to expand $N_x^\mu[h]$ and $Q_{x,x'}^{\mu\nu}[h]$ to respectively first and zeroth order in $h$ to get $\Wc[h]$ to second order in $h$. In particular, $Q_{x,x'}^{\mu\nu}[h=0]$, i.e. $Q^{\mu\nu}$ evaluated in flat space, has been computed above and is given by \cref{eq:defQbubble}. The action \labelcref{eq:inducedActionFormal} thus simplifies to
\begin{align}
\Wc[A,h]={}&\Wc_0[h] + \Wc_\theta[h]  \label{eq:indActFormalDecomp}
\end{align}
where we introduced
\begin{equation}
\Wc_\theta[h]  = -\dfrac{1}{2}\int_{q} N^{\mu}_{-q}[h] \dfrac{q_\mu q_\nu}{q_\alpha q_\beta Q^{\alpha \beta}(q)} N^{\nu}_q[h].
\end{equation}

The two terms on the rhs of \cref{eq:indActFormalDecomp} have different origins. $\Wc_0[h]$ is the contribution to the induced action one would get at the mean-field level; i.e., by discarding the phase mode $\theta$, and we dub it the \emph{pure geometric} contribution. On the other hand, we call $\Wc_\theta[h]$ the \emph{phase} contribution, as it corresponds to what is obtained by integrating out the phase mode, with two vertices $q_\mu N^{\mu}_q[h]$ representing an effective interaction between the metric $h$ and the phase mode $\theta$ linked by the inverse Goldstone propagator $q_\alpha q_\beta Q^{\alpha \beta}(q)$.

 Since we are interested only in determining $\eta^{(1)}_\orm$ and $\eta^{(2)}_\orm$ at leading order in momentum and frequency, we organize the computation accordingly.  Dimensional analysis suggests that, at leading order, $\eta^{(1)}_\orm = \Oc(|\qv|^0)$ and $\eta^{(2)}_\orm = \Oc((c_s^{-2}\omega^2-\qv^2)^{-1})$, an intuition confirmed by the effective field theory calculation [see \cref{EFT:eta12cs}, and \cite{Golan2019} for the calculation up to the next-to-leading order]. Either  term $\Wc_0[h]$ and $\Wc_\theta[h]$ brings a different contribution to the viscosity tensor. The one-loop integrals appearing in $\Wc_0[h]$ and the vertices $q_\mu N^{\mu}_q[h]$ are all regular in the infrared limit $\qv \to 0$, $\omega \to 0$ and thus, the only way a term of order $\Oc(|\qv|^{-2})$ can appear in the calculation is through the inverse Goldstone propagator $\sim (c_s^{-2}\omega^2-\qv^2)$ from the phase contribution.
 Hence, the leading contribution to $\eta^{(2)}_\orm$ is entirely fixed by $\Wc_\theta[h]$. Conversely, as the Goldstone propagator does not appear in $\eta^{(1)}_\orm$ at the leading order, $\eta^{(1)}_\orm$ is determined by $\Wc_0[h]$.
We now compute each contribution  to the viscosity separately.

\paragraph{Contribution from the pure geometric part}

\begin{figure}
\centering
    \begin{subfigure}[m]{0.4\textwidth}
    \centering 
    \includegraphics{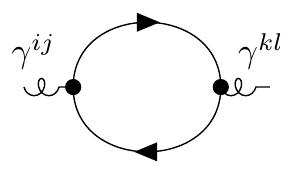}
        \caption{Bubble contribution to $\Wc_0[h]$.}
        \label{fig:Eta1}
    \end{subfigure}
~
    \begin{subfigure}[m]{0.4\textwidth}
    \centering 
    \includegraphics{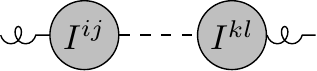}
        \caption{Phase mode contribution to $\Wc_\theta[h]$.}
        \label{fig:Eta2}
    \end{subfigure}
    \par
\caption{Diagrammatic representation of the loop integrals contributions to the pure geometric (a) and phase parts (b) of the induced action. The curly line stands for the metric $h_{ij}$ insertions, the dots for the fermion-metric interaction vertices $\gamma^{ij}$ [\cref{eq:defVertGam}] and the grey blobs for the phase-metric interaction vertices $I^{ij}$ [\cref{eq:exprIqN}].}
\end{figure}

Here we start from the fermionic action defined by the inverse propagator \labelcref{eq:FullNambuProp}, discarding for now the phase mode $\theta$. We work at vanishing $\Uone_N$ gauge field, i.e. $\Ac = 0$. The fermions are integrated out following \cref{eq:loopExpansionSeff}, where $\Gamma$ is given by \cref{eq:defGammaGeneral} with $\Ac$ set to zero. The detailed calculation is presented in \cref{app:MicCalcCSnoC}. The end result is
\begin{equation}
\Wc_0[h] = \dfrac{1}{2} \int_q h_{-q,ij}  R^{ij,kl}(q) h_{q,kl}
\label{eq:KernelRac}
\end{equation}
 with
\begin{align}
R^{ij,kl}(q)&= \dfrac{1}{2} \int_p \tr [\Gc_{0,p} \gamma^{ij}_{p,-p-q} \Gc_{0,p+q} \gamma^{kl}_{p+q,-p}],
\label{eq:KernelR}
\\
\gamma_{q,-q'}^{ij} {}&{}= \dfrac{1}{2}  \delta^{ij}
\bigg(
(2q_0-q'_0) \sigma^0 - \dfrac{\qv\cdot\qv'}{2m} \sigma^z + \Delta q'_k \sigt^k
\bigg)-  \dfrac{q^{(i} q'{}^{j)}}{2m} \sigma^z + \dfrac{1}{2} \Delta q'{}^{(i} \sigt^{j)}.
\label{eq:defVertGam}
\end{align}
The tensor $R^{ij,kl}(q)$ is defined by a one-loop integral analogous to the polarization bubble $Q^{ij}$ [\cref{eq:defQbubble}] that is relevant in the calculation of the conductivity. It is represented diagramatically in \cref{fig:Eta1}.
 From \cref{eq:relWsig}, it is related to the odd viscosity through
\begin{equation}
\eta^{ijkl}_\orm(\omega,\qv) = \dfrac{4}{\ir \omega^+} R^{\Rrm,ijkl}(\omega,\qv),
\end{equation}
where $R^{\Rrm,ijkl}(\omega,\qv)=R^{\Rrm,ijkl}(\ir \omega \to \omega^+,\qv)$.
  The odd viscosities are obtained by projecting onto the odd tensors $\sigma^{xz}$ for $\eta^{(1)}_\orm$ and $\sigma^{0x}$ or $\sigma^{0z}$ for  $\eta^{(2)}_\orm$, as done in \cref{app:MicCalcEtaGeonoC}. At small frequency and momentum, one gets
 \begin{align}
 \eta_{\orm}^{(1)}(\omega,\qv^2) &= \dfrac{s n_0}{2} + \Oc(\omega^2,\qv^2),
\label{eq:exprEta1MicroNoC}
\end{align} 
while the contribution to $\eta_{\orm}^{(2)}$ is of order $\Oc(|\qv|^0)$, i.e. is subleading.

\paragraph{Contribution from the phase}

We now compute the contributions to the viscosity tensor that originate from the phase mode. The linear term $N^{\mu}_q[h]$ is determined to leading order in $h_{ij}$ in \cref{app:MicCalcPhaseContribnoC}, yielding
\begin{gather}
q_\mu N^{\mu}_q[h] {}= h_{ij,q} I^{ij}_q, \label{eq:defIqN}\\
 I^{ij}_q  {}= -\dfrac{1}{2} \tr \int_p \Gc_{0,p}\bigg[q_0 \sigma^z +\dfrac{(2 \pv + \qv)\cdot \qv}{2m} \sigma^0\bigg] \Gc_{0,p+q}
\bigg[
\dfrac{(p+q)^{(i}p^{j)}}{2m}\sigma^z +\dfrac{1}{2} \Delta \sigt^{(i}p^{j)}
\bigg].\label{eq:exprIqN}
\end{gather}
The resulting contribution to the viscosity tensor $\eta^{ijkl}$ is given by
\begin{equation}
\eta^{ijkl}_\orm(\omega,\qv) = \dfrac{4}{\ir \omega^+} S^{\Rrm,ijkl}(\omega,\qv)
\label{eq:relEtaS}
\end{equation}
with $S^{\Rrm,ijkl}(\omega,\qv)$ the retarded part of 
\begin{equation}
S^{ijkl}(q) = -\dfrac{I^{ij}_{-q}I^{kl}_q}{q_\alpha q_\beta Q^{\alpha \beta}(q)},
\label{eq:defSijkl}
\end{equation}
represented diagramatically in \cref{fig:Eta2}.

The projection of the corresponding viscosity tensor on $\sigma^{ab}$ matrices is done in \cref{app:MicCalcPhaseContribnoC}. For $\eta_\orm^{(1)}$, the projection vanishes, hence the contribution  from the phase mode to $\eta^{(1)}_\orm$ is at most of order $\Oc(|\qv|^4/(c_s^{-2}\omega^2-\qv^2))$. For $\eta_\orm^{(2)}$, the direct calculation yields 
\begin{equation}
\eta_\orm^{(2)}(\omega,\qv^2)=- \dfrac{s n_0}{2} \dfrac{ c_s^{2}}{\omega^2- c_s^{2}\qv^2}.
\label{eq:exprEta2MicroNoC}
\end{equation}

Having determined the  odd conductivity \labelcref{eq:exprSigHMicroNoC} as well as the two components of the odd viscosity tensor, \labelcref{eq:exprEta1MicroNoC,eq:exprEta2MicroNoC}, one checks that the Ward identity \labelcref{WardH} is satisfied. 
  Contrary to the case of the conductivity, where the incorporation of the phase mode is crucial to preserve the $\Uone_N$  gauge invariance and get the correct result for the associated transport coefficient $\sigma^{\Hrm}$, it is not obvious \emph{a priori} whether including the phase mode is important or not to obtain the viscosity tensor. This is reflected in the calculation as the phase mode   
  does not affect the value of $\eta^{(1)}_\orm$ but is crucial to obtain $\eta^{(2)}_\orm$. We notice that only by going beyond mean-field is the Ward identity  \labelcref{WardH} fulfilled, as expected since the mean-field theory breaks down $\Uone_N$  gauge invariance which the identity relies on.

\subsection{Inclusion of Coulomb and non-local interactions}
\label{sec:sub_Coul}

In this section, we now additionally incorporate a non-local Coulomb interaction between the fermions, for which the corresponding Euclidean action reads in flat space
\begin{equation}
S_{\Crm,\frm}[\psi,\psi^*] = \dfrac{1}{2} \int_{x,x'} (\psi^\dagger\psi - \bar{n})V(\rv-\rv')(\psi^\dagger\psi - \bar{n})
\label{eq:act_coulomb1}
\end{equation}
with $V(\rv)$ the interaction potential, $e$ the electric charge and $\bar{n}$ the background density. 
We consider potentials  satisfying $V(\qv)\sim |\qv|^{-\alpha}$ at long distances ($\qv \to 0$), with $0\le \alpha <2$. The case $\alpha=0$  corresponds to  of short-ranged interactions, $\alpha=1$ to a mixed-dimensional Coulomb interaction, and $\alpha=2$ to an intrictically two-dimensional Coulomb potential, see  \cref{sec:EFTmixedSC} for a more thourough discussion.

The density-density interaction in \cref{eq:act_coulomb1} is decoupled by means of a Hubbard-Stratanovitch transform, with an auxiliary field $\chi$, yielding
\begin{equation}
S_{\Crm,\frm}[\psi,\psi^*,\chi] = \dfrac{1}{2}\int_x \chi V^{-1} \chi   + \int_x (\psi^*\psi-\bar{n}) (-\ir \chi),
\label{eq:act_coulombHSflat}
\end{equation}
where $V^{-1} = (-\nablav^2)^{\alpha/2}/e^2$ is the inverse propagator for the Coulomb field. The expression \labelcref{eq:act_coulombHSflat} allows to generalize the theory to arbitrary curved space\footnote{To formulate the theory in curved space starting from the action \labelcref{eq:act_coulomb1}, one would need to replace the distance $\rv-\rv'$ by the geodesic distance \cite{Bradlyn2012}}
\begin{equation}
S_{\Crm}[\psi,\psi^*,\chi;g] = \dfrac{1}{2}\int_x \sqrt{g} \chi V^{-1} \chi   + \int_x \sqrt{g} (\psi^*\psi-\bar{n}) (-\ir \chi)
\label{eq:act_coulombHScurved}
\end{equation}
with the Laplace operator in $V^{-1}$ replaced by the covariant Laplacian $g^{ij}\nabla_i \nabla_j$.

\begin{figure}
{\centering
$
\vcenter{\hbox{\includegraphics{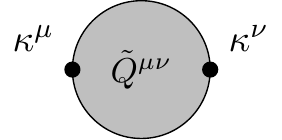}}}
=
\vcenter{\hbox{\includegraphics{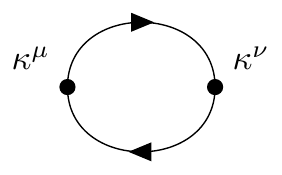}}}
+
\vcenter{\hbox{\includegraphics{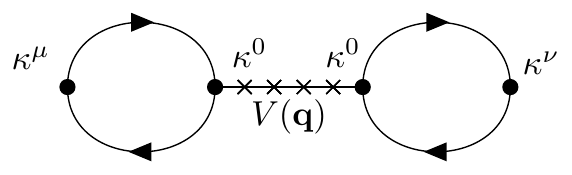}}}
{}+{}\dotsb
$\par
}
\caption{Renormalization of the polarization bubbles $Q^{\mu\nu}(q)$ due to the Coulomb interaction.  The crossed line denotes the Coulomb potential $V(\qv)$ which only couples to fermions through the density vertices $\kappa^0$.
\label{fig:Qt}
}
\end{figure}

The Coulomb interaction couples to the fermions like the time-component of the $\Uone_N$ gauge field. Hence, the total action $S_\Trm=S+S_\Crm$ reads
\begin{equation}
S_\Trm [\Psi,\Psi^\dagger;\Act,h] = S[\Psi,\Psi^\dagger;\Act,h] +  \dfrac{1}{2}\int_x \sqrt{g} \chi V^{-1} \chi + \ir \bar{n}\int_x \sqrt{g} \chi,
\end{equation}
where $S$ is the action \labelcref{eq:effactForm} and  we introduced $\Act_0=\Act_0+\chi$, $\Act_i=\Ac_i$. We integrate out the fermions following \cref{eq:loopExpansionSeff,eq:formalExpansionSeff} to obtain the effective action for $\phi$, $\chi$, $A$,  and $h$ to quadratic order in fields and sources,
\begin{align}
S_\Trm[\Act,h]={}&\Wc_0[h] + \int_x \Act_{\mu,x} N^{\mu}_x[h]  + \dfrac{1}{2} \int_{x,x'} \Act_{\mu,x} Q^{\mu\nu}_{x,x'}[h]\Act_{\nu,x'} \nnl
{}&{}+ \ir \bar{n}\int_x \sqrt{g} \chi +  \dfrac{1}{2}\int_x \sqrt{g} \chi V^{-1} \chi.
\label{eq:act_coulInterm}
\end{align}
At this stage, we integrate out the Coulomb interaction. The linear terms in $\chi$ proportional to $N^{\mu}_x[h]$ and to $\bar{n}$ in \cref{eq:act_coulInterm} compensate each other to ensure charge neutrality and the effective action for $\Ac$ and $h$ reads
\begin{equation}
S[\Ac,h]=\Wc_0[h] + \int_x \Ac_{\mu,x} N^{\mu}_x[h] + \dfrac{1}{2} \int_{x,x'} \Ac_{\mu,x} \Qt^{\mu\nu}_{x,x'}[h]\Ac_{\nu,x'},
 \label{eq:formalExpansionSeff_COUL}
\end{equation}
where $\Qt[h]$ is defined by
\begin{equation}
\Qt_{\mu\nu}[h] = Q_{\mu\nu}[h] + Q_{\mu 0}[h]\dfrac{V}{\sqrt{g}-V Q_{00}} Q_{0\nu}[h]
\label{eq:defQtcurved}
\end{equation}
with the products defined in the operator sense. In the specific case of flat space, $\Qt_{\mu\nu}=\Qt_{\mu\nu}[h=0]$ reduces to
\begin{gather}
\begin{aligned}
\Qt_{00}(q)&=\dfrac{Q_{00}(q)}{1-V(\qv) Q_{00}(q)},&
\Qt_{j0}(q)&=\dfrac{Q_{j0}(q)}{1-V(\qv) Q_{00}(q)},
\end{aligned}
\label{eq:defQtflat1}
\\
\Qt_{ij}(q)=Q_{ij}(q)+\dfrac{V(\qv) Q_{i 0}(q)Q_{0 j}(q)}{1-V(\qv) Q_{00}(q)}.
\label{eq:defQtflat2}
\end{gather}
Expanding the Coulomb field to quadratic order and integrating it out is equivalent to the random phase approximation (RPA) and expressions \labelcref{eq:defQtcurved,eq:defQtflat1,eq:defQtflat2} can equivalently be obtained through resummation of the bubble diagrams illustrated in \cref{fig:Qt}.

 Comparing the actions \labelcref{eq:formalExpansionSeff} and \labelcref{eq:formalExpansionSeff_COUL}, the only difference is that the kernel $Q_{\mu\nu}[h]$ has been replaced by $\Qt_{\mu\nu}[h]$. It is straightforward now to extract the Hall transport coefficients. Following \cref{sec:sub_condNoCoul}, one finds the induced action for the $\Uone_N$ gauge field in flat space to be
\begin{align}
\Wc[A] &= 
\dfrac{1}{2}\int_q A_{-q,\mu} \Kt^{\mu\nu}(q) A_{q,\nu},&
\Kt^{\mu\nu}(q) &= \Qt^{\mu\nu}(q) -  \dfrac{\Qt^{\mu\rho}(q)q_\rho q_\sigma \Qt^{\sigma\nu}(q)}{q_\alpha q_\beta \Qt^{\alpha\beta}(q)},
\end{align}
yielding the Hall conductivity
 \begin{equation}
\sigma_\Hrm (\omega,\qv^2)=\dfrac{s n_0}{2 m^2} \dfrac{-\qv^2}{\omega^2 -  \omega_p^2|\qv|^{2-\alpha}- c_\srm^2 \qv^2}.
 \end{equation}
 As for the viscosity, we follow \cref{sec:sub_viscNoCoul}. The calculation of $\eta_\orm^{(1)}$ does not involve $\Qt$ and $\eta_\orm^{(1)}$ is thus unaffected by the interaction. On the other hand, $\eta_\orm^{(2)}$ is
 related to 
\begin{equation}
\tilde{S}^{ijkl}(q) = -\dfrac{I^{ij}_{-q}I^{kl}_q}{q_\alpha q_\beta \Qt^{\alpha \beta}(q)},
\end{equation}
through \cref{eq:relEtaS}, and thus becomes
 \begin{equation}
 \eta_\orm^{(2)}(\omega,\qv^2)= - \dfrac{s n_0}{2} \dfrac{1}{|\qv|^\alpha} \dfrac{c_s^2 |\qv|^\alpha + \omega_p^2}{\omega^2-  \omega_p^2 |\qv|^{2-\alpha}-c_s^2\qv^2}
 \end{equation}
 in presence of the long-range interaction $V(\rv)\sim |\rv|^{2-\alpha}$.
 
 The Hall conductivity and viscosities obtained here agree with what was found in \cref{sec:EFTintrinsticSC,sec:EFTmixedSC} from the effective field theory approach.

\section{Discussion and outlook} \label{sec:conc}
In this paper we computed electromagnetic and geometric linear responses in non-relativistic two-dimensional chiral superconductors, where in addition to short-range attractive interactions that lead to chiral pairing, elementary fermions interact via an instantaneous long-range Coulomb potential. For the two-dimensional logarithmic Coulomb interaction we found that the homogeneous $\mathbf{q}\to 0$ limit of the Hall viscosity tensor $\eta_{\mathrm{o}}$ is ill-defined because the result depends on the direction of the momentum vector $\mathbf{q}$. We believe that this peculiar behavior is an artifact of the instantaneous nature of the Coulomb potential. It is expected that the problematic  $\mathbf{q}^{2}$ denominator of the component $ \eta_{\mathrm{o}}^{(2)}$ in \cref{EFT:eta12csSC} is replaced by $\mathbf{q}^2-\omega^2/c^2$ after  the Coulomb potential is replaced by a retarded electromagnetic interaction that propagates with a finite speed of light $c$. In this way at a finite frequency $\omega$ the $1/\mathbf{q}^{2}$ singularity will be regularized. To clarify this issue in a future work we are planning to compute the Hall viscosity and conductivity in the Lorentz-invariant chirally paired model \cite{Golkar2014} coupled to the Maxwell electromagnetism. This approach can also shed new light on the geometric Meissner effect \cite{PhysRevLett.120.217002} and the geometric induction \cite{jiang2019geometric} in chiral superconductors.

It would be interesting to extend the ideas presented in this paper to non-abelian quantum Hall states (such as Pfaffian, anti-Pfaffian, particle-hole symmetric Pfaffian) which can be viewed as chiral paired states of composite fermions \cite{Read2000, Son2015} that couple to a dynamical $2+1$ dimensional abelian gauge field.

Inspired by recent work on viscoelastic linear response of anisotropic systems \cite{souslov2019anisotropic, rao2019hall}, it would be interesting and straightforward to extend this work to anisotropic chiral superconductors.

\section*{Acknowledgements}
We thank Ady Stern and Carlos Hoyos for productive discussions and comments on the manuscript.
We acknowledge fruitful discussions with Nicolas Dupuis, Dam Thanh Son, Anton Souslov and Wilhelm Zwerger.  OG acknowledges support from the Israel Science Foundation
(ISF), the Deutsche Forschungsgemeinschaft
(DFG, CRC/Transregio 183, EI 519/7-1), and the European Research Council (ERC, Project LEGOTOP). The work of SM is funded by the Deutsche Forschungsgemeinschaft (DFG, German Research Foundation) under Emmy Noether Programme grant no.~MO 3013/1-1 and under Germany's Excellence Strategy - EXC-2111 - 390814868. 

\begin{appendix}

\section{Ward identities}
\label{app:ward}

In this Appendix we derive the continuity equations \labelcref{eq:continuityCurrent,eq:continuityTens} and the Ward identity \labelcref{eq:WardMaster}. Our starting point is \cref{eq:invarEffAct} which implies that for the infinitesimal transforms defined by \labelcref{eq:transfLawAt,eq:transfLawAi,eq:transfLawg}
\begin{align}
\int_x \delta A_\mu(x) \Wc^{(1,0)\mu}[x;A,g] + \delta g_{ij}(x) \Wc^{(0,1)ij}[x;A,g]=0.
\end{align}
This holds for any choice of $\alpha(x)$, $\xi^i(x)$ and thus
\begin{align}
\partial_\mu \Wc^{(1,0)\mu}&=0,	\label{eq:AppWWardGauge}\\
m \partial_t (\Wc^{(1,0)}_k) - [(\partial_j g_{ik} + \partial_i g_{jk}-\partial_k g_{ij})\Wc^{(0,1)ij} {}&{}+ (g_{ik}\partial_j\Wc^{(0,1)ij} + g_{jk}\partial_i\Wc^{(0,1)ij})]\nnl
&= (\partial_\mu A_k-\partial_k A_\mu) \Wc^{(1,0)\mu}			,\label{eq:AppWWardmetric}
\end{align} 
where every quantity is evaluated at arbitrary spacetime point and for any value of the sources and $k$ is an arbitrary space index. The indices are raised and lowered using the spatial metric tensor $g$.
 Using now \cref{eq:CurrentExpVal} one gets the continuity equations, \cref{eq:continuityCurrent,eq:continuityTens}.

By taking a further derivative of \cref{eq:AppWWardmetric} wrt $A_\nu$ and setting $A=0$, $g=\delta$ we get
 in Fourier space
\begin{gather}
- m \omega^+ \Wc^{(2,0)k\nu}(q)-[q_j\Wc^{(1,1)\nu,kj}(-q) + q_i\Wc^{(1,1)\nu,ik}(-q)] \nnl
+ \dfrac{1}{\sqrt{V}}\{
q^k\Wc^{(1,0)\nu}(q=0) + \delta^k_\nu [\omega^+\Wc^{(1,0)t}(q=0)-q_i\Wc^{(1,0)i}(q=0)]
\} =0. \label{eq:AppWWardmetric10}
\end{gather}
We also differentiate \cref{eq:AppWWardmetric} wrt $g_{ab}$.
 This gives\footnote{To get a fully symmetric expression we remark that $\Wc^{(0,1)ij}=\Wc^{(0,1)(ij)}=(\Wc^{(0,1)ij}+\Wc^{(0,1)ji})/2$ and use $\delta g^{ab}/\delta g^{ij}=(\delta^a_i\delta^b_j+\delta^a_j\delta^b_i)/2$.}
 \begin{gather}
 -m \omega^+ \Wc^{(1,1)k,ab}(q) -	2q_j \Wc^{(0,2)(kj),ab}(q)\nnl
 {}= \dfrac{1}{\sqrt{V}}\big\{\delta^{kb}q_j \Wc^{(0,1)(aj)} (q=0) + \delta^{ka}q_j \Wc^{(0,1)(bj)} (q=0) - q^k \Wc^{(0,1)ab} (q=0)\big\} 
. \label{eq:AppWWardmetric01}
 \end{gather}
We contract \cref{eq:AppWWardmetric01} with $q_a$ and cancel the resulting $q_a\Wc^{(1,1)k,ab}$ term with the one appearing in  \cref{eq:AppWWardmetric10} to get
\begin{gather}
m^2 (\omega^+)^2 \Wc^{(2,0)ij}(q)= 4q_kq_l\Wc^{(0,2)ikjl}(q)+\dfrac{1}{\sqrt{V}}\{2\delta^{ij} q_k q_l\Wc^{(0,1)kl}(q=0)\nnl
 -m\omega^+ [q^i\Wc^{(1,0)j}(q=0)+\delta^{ij}(\omega^+\Wc^{(1,0)t}(q=0)-q_k\Wc^{(1,0)k}(q=0))]\}. \label{eq:AppWwardCondVisc}
\end{gather}
The second derivatives of the induced action $\Wc$ can be related to physical observables via \cref{eq:relWsig,eq:relWeta}.  
For vanishing sources
\begin{align}
\braket{J^{\mu}}&=0,&
\braket{T^{ij}}&=P \delta^{ij}
\end{align}
and to leading order in momentum
\begin{equation}
\lambda^{ijkl}(\qv,\omega) = P(\delta^{ik}\delta^{jl} +\delta^{il}\delta^{jk})+\kappa^{-1}\delta^{ij}\delta^{kl} + \Oc(|\qv|)
\end{equation}
with $P$ the pressure and $\kappa^{-1}=-V \big(\partial P/ \partial V \big)_{S, N}$ the inverse compressibility. Using these results in \cref{eq:AppWwardCondVisc} yields the Ward identity \labelcref{eq:WardMaster}.

\section{Gaussian functional integration in curved space}
\label{app:gaus}
Here we provide some details on how Gaussian functional integration is performed in the presence of a general background metric $g_{ij}(t,\mathbf{x})$.

Consider a $d$-dimensional Gaussian scalar theory defined in a general coordinate system by the Euclidean action
\beq
S_\Erm=\int_x \sqrt{g} \bigg[\frac 1 2 \varphi a \varphi+ b \varphi \bigg],
\eeq
where $g=\det g_{ij}$, $a$ is a symmetric operator and $b$ is some given scalar function of space and time.

The general-coordinate invariant inner product in the space of scalar fields is given by
\beq
\langle \varphi_1| \varphi_2 \rangle= \int_x \sqrt{g} \varphi_1 \varphi_2,
\eeq
where the volume element  $\sqrt{g(t,\mathbf{x})}$ can be viewed as a (diagonal) metric tensor in the space of scalar fields. It follows that the general-coordinate invariant functional measure is \cite{hawking1977zeta, PhysRevLett.42.1195, PhysRevD.35.3796}
\beq
\mathcal{D}[\varphi]= \prod_{\mathbf{x}} g^{1/4}(\mathbf{x}) \dr \varphi(\mathbf{x}).
\eeq
As a result, the Gaussian functional integral in Euclidean time can be performed as follows,
\begin{align}
\int \mathcal{D}[\varphi] \er^{-\int_x \sqrt{g} \big[\frac 1 2 \varphi a \varphi+ b \varphi \big]} 
{}={}&\frac{\prod_{\mathbf{x}} g^{1/4}}{\sqrt{\det{\left(\sqrt{g} a/2\pi \right) }}} \er^{\frac 1 2 \int_x  \big[ \sqrt{g}b \left(\sqrt{g} a \right) ^{-1} \sqrt{g} b \big] }\nnl
{}={}&\frac{1}{\sqrt{\det (a/2\pi)}} \er^{\frac 1 2 \int_x  \sqrt{g}   b a^{-1} b  }.
\end{align}

\section{Linear response from EFT of two-dimensional chiral superconductors}
\label{app:LR}

In this Appendix we compute to leading order in derivatives and
within RPA the induced action resulting from the EFT \labelcref{EFT:acscm}
which describes a chiral superconductor, where charged fermions interact
via a two-body potential whose form in Fourier space is $V(\qv)\sim|\mathbf{q}|^{-\alpha}$.
We consider the range $0\leq\alpha\leq2$, which includes short range interactions
at $\alpha=0$, the 3$d$ Coulomb interaction $V(\rv)\sim1/|\mathbf{r}|$
at $\alpha=1$, and the 2$d$ Coulomb interaction $V(\rv)\sim\log|\mathbf{r}|$
at $\alpha=2$.

 In \cref{app:legendre} we do so using a formalism based on the Legendre transform of the pressure functional. That calculation, complementary to what is done in \cref{sec:EFTmixedSC,sec:EFTintrinsticSC}, unifies the superconductor EFTs \labelcref{EFT:acscm,EFT:acsc} with the superfluid EFT
\labelcref{EFT:ac} by introducing the renormalized speed of sound.
We then derive and analyze the resulting linear response functions in \cref{subsec:Induced-action-and}.

\subsection{Legendre transform interpretation of the effective theory and renormalized
speed of sound}
\label{app:legendre}

\subsubsection{Generalities }

We start by generalizing \cref{EFT:acscm} to

\begin{align}
S =\int_x \sqrt{g}\bigg[P (X-\chi ) + n_{0}\chi+\frac{1}{2}\chi\frac{1}{V(-\nablav^{2})}\chi\bigg],\label{eq:63-1-3}
\end{align}
where $V(\qv^2)$ is the Fourier transform of a density-density
interaction between microscopic fermions. We then rewrite $P(X)=X\rho-\varepsilon\left(\rho\right)$,
where $\rho$ is a new dynamical field which physically corresponds
to the density, and $\varepsilon=\mathcal{L}\{ P\} $ is
the Legendre transform of $P$, the internal energy density of the
superfluid \cite{Hoyos2013}. \cref{eq:63-1-3} can then
be written as 
\begin{align}
S & =\int_x\sqrt{g}\bigg[X\rho-\varepsilon (\rho)-(\rho-n_{0})\chi+\frac{1}{2}\chi\frac{1}{V(-\nablav^{2})}\chi\bigg].\label{eq:63-1-1-2}
\end{align}
Since $\chi$ appears quadratically it can be integrated out exactly,
leading to 
\begin{align}
S & =\int_x \sqrt{g}(\rho X-\tilde{\varepsilon}[\rho]),\label{eq:63-1-1-1-2}
\end{align}
where $\tilde{\varepsilon}[\rho]=\varepsilon(\rho)-(\rho-n_{0})V(-\nablav^{2})(\rho-n_{0})/2$
is the internal energy density of the superfluid, supplemented by
the interaction energy. The square brackets indicate that $\tilde{\varepsilon}$
is generally a functional, as opposed to the function $\varepsilon$.
Integrating out $\rho$ is now equivalent to a (functional) Legendre
transform from $\tilde{\varepsilon}$ to $\tilde{P}=\mathcal{L}\{ \tilde{\varepsilon}\}$,\footnote{Note that the Legendre transform is an involution, $\mathcal{L}^{-1}=\mathcal{L}$.}
\begin{align}
S & =\int_x\sqrt{g}\tilde{P}[X].\label{eq:63-1-1-1-1-1}
\end{align}
The superconductor EFT \labelcref{EFT:acscm} is therefore equivalent to the superfluid
EFT \labelcref{EFT:ac}, with a renormalized pressure functional $\tilde{P}[X]$ which simply
accounts for the additional interaction energy $V(\qv^2)$ of microscopic
fermions.

\subsubsection{Quadratic pressure functional and speed of sound}

Let us now assume that the pressure functional $P(X)$ is quadratic,
\begin{align}
 P(X)=P_{0}+n_{0}(X-\mu)+\frac{1}{2}\frac{n_{0}}{mc_{s}^{2}}(X-\mu)^{2}.\label{eq:66}
\end{align}
In this case the Legendre transform $\tilde{P}[X]$ can be computed
explicitly  and one finds 
\begin{align}
\tilde{P}[X]= & P_{0}+n_{0}(X-\mu)+\frac{1}{2}\frac{n_{0}}{mc_{s}^{2}}(X-\mu)\frac{1}{1+n_{0}V(-\nablav^{2})/mc_{s}^{2}}(X-\mu).
\end{align}
Comparing with \cref{eq:66}, we see that the only effect of the
interaction $V(\qv^2)$ is to renormalize the speed of sound, $\cs^{2}\to\cst^{2}=c_{s}^{2}+n_{0}V(-\nablav^{2})/m$,
where $\cst$ has been promoted from a number to an operator. This reads  in Fourier space 
\begin{align}
\cs^{2}\to\cst^{2}(\qv^{2})= & \cs^{2}+\omega_{p}^{2}|\qv|^{-\alpha}\label{eq:68}
\end{align}
for $V(\qv^{2})=e^{2}|\qv|^{-\alpha}$. Thus, for quadratic $P$,
the superconductor EFT is identical to the superfluid EFT with a
renormalized speed of sound. Note that this renormalization is trivial
for the contact interaction $\alpha=0$, since in this case $\cs$
is independent of $\qv$, as in the superfluid. Using the thermodynamic
expression $K^{-1}=n^{-1}(\delta n/\delta\tilde{P})_{T}=n_{0}m\cs^{2}$,
\cref{eq:68} implies the (zero temperature) inverse compressibility
\begin{align}
K^{-1}(\qv)= & n_{0}m(c_{s}^{2}+\omega_{p}^{2}|\qv|^{-\alpha}).\label{eq:69}
\end{align}
In particular, for all $\alpha>0$ the superconductor is incompressible
with respect to a uniform compression, $K(q=0)=0$. In
 \cref{subsec:Induced-action-and} we will use \cref{eq:68}
to deduce the induced action of the superconductor from that of the
superfluid.

\subsubsection{Approximations: linear response, derivative expansion, and RPA \label{app:approximations}}

The pressure functional $P$ is not quadratic in general, and in particular in
the microscopic model of \cref{sec:microTheory}. However, for the purpose of computing
linear response functions, to lowest order in derivatives, and within
RPA, only an expansion of $P$ to quadratic order \labelcref{eq:66} around $\mu$
is sufficient, because the effective theory \cref{eq:63-1-3} can be
expanded to second order in all fields. Indeed, linear response is extracted from the quadratic expansion in background fields. To lowest order in derivatives,
diagrams with the Goldstone field $\varphi$ running in loops can
be neglected \cite{Son2006,Golan2019}, which amounts
to a quadratic expansion in $\varphi$. Finally, the RPA amounts to
neglecting diagrams with $\chi$ running in loops, i.e a quadratic
expansion in $\chi$. Explicitly, to second order in all fields \cref{eq:63-1-3}
reduces to 
\begin{align}
S=\int_x\sqrt{g}  \bigg\{ & P_{0}-n_{0}D_{t}\varphi+\frac{n_{0}}{2m}[c_{s}^{-2}(D_{t}\varphi)^{2}-(\Dv \varphi)^{2}]\nnl
 &{} +\frac{n_{0}}{m \cs^{2}}\bigg[\chi D_{t}\varphi+\frac{1}{2}\chi(c_{s}^{2}\omega_{p}^{-2}(-\nablav^{2})^{\alpha}+1)\chi\bigg]\bigg\}, 
\end{align}
and integrating over $\chi$ leads to the speed of sound renormalization
\begin{align}
S & =\int_x \sqrt{g}\bigg\{ P_{0}-n_{0}D_{t}\varphi+\frac{n_{0}}{2m}[\tilde{c}_{s}^{-2}(D_{t}\varphi)^{2}-(\Dv\varphi)^{2}]\bigg\} \label{eq:70}
\end{align}
which is nothing but the action \labelcref{eq:lagPhiEftCoul}.
We note that space and time derivatives are counted equally, and that
the leading derivative corrections to \labelcref{eq:70} are due to second
order terms that add to $P\left(X\right)$ in the superfluid EFT \cite{Son2006,Golan2019}, rather than $\varphi$ loops. As a result,
the linear response functions obtained in \cref{subsec:Induced-action-and}
below, which can formally be expanded to infinite order in derivatives,
receive corrections at order $L+2$, where $L$ is their leading order
in derivatives. In particular,  $\sigma_\text{H}$ in \cref{Hallcon} below, is of order $\alpha$, and receives derivative corrections at order $\alpha+2$. Additionally, $\eta_{\text{o}}^{(2)}$ in \cref{Hallvis} is of leading order $-2$, and receives corrections at zeroth order.  

\subsection{Induced action and linear response\label{subsec:Induced-action-and}}

Since the quadratic effective action \labelcref{eq:70} is identical to that
of the superfluid under the replacement $c_{s}\to\tilde{c}_{s}$,
the quadratic induced action in the superconductor can be directly
obtained from that of the superfluid by applying the same replacement.
The induced action in chiral superfluids was computed and analyzed
in Refs.\cite{Hoyos2013,Golan2019}, and below we use the results of
Appendix D of Ref.\cite{Golan2019}. Replacing $\cs\to\cst$, the induced action of the chiral superconductor expanded around flat space is
we obtain 
\begin{align}
\Wc[g,A]= & \int_x \Big[2 P_{0} h-n_{0}\mathsf{A}_{t} \nnl {}&{}+\dfrac{1}{2}\dfrac{n_{0}}{m}\dfrac{\cst^{2}\Bsf^{2}-\Esfv^{2}+(\ir s/m)\mathsf{E}^{i}q_{i}B-(s^{2}/4m^{2})\qv^{2}B^{2}}{\omega^{2}-\cst\qv^{2}}\nonumber \\
 & +2n_{0}\dfrac{\tilde{c}_{s}^{2}h[\ir q_{i}\mathsf{E}^{i}+(s/2m)\qv^{2}B]-m\cst^{2}\omega^{2}h^{2}}{\omega^{2}-\cst^{2}\qv^{2}}\Big],\label{eq:71} 
\end{align}
where $h_{ij}=g_{ij}-\delta_{ij}$ and $h= h^i_i$.
This expression encodes the entire linear response of the chiral superconductor in flat space  within RPA
 and to the lowest order in derivatives, as we now discuss.

First, note that the terms in $\Wc$, and the corresponding linear response
functions, can be split into three groups, according to their dependence
on $\tilde{c}_{s}$. The terms in the first group are independent
of $\tilde{c}_{s}$, and accordingly, they are unaffected by the additional fermion
interaction $V$ that distinguishes the superconductor from the superfluid.
These appear in the first line of \labelcref{eq:71}, and include the
ground state pressure $P_{0}$ and density $n_{0}$, as well the first
odd viscosity $\eta_{\text{o}}^{\left(1\right)}=sn_{0}/2$ [\cref{EFT:eta12csSCmixed}].

The second group of terms in $\Wc$ includes those where $\tilde{c}_{s}$
appears only in the denominator, and include the terms $-\Ev^{2}+(\ir s/m)E^{i}q_{i}B$ be derived from $-\boldsymbol{\mathsf{E}}^{2}+(\ir s/m)\mathsf{E}^{i}q_{i}B$
in the second line of \labelcref{eq:71}, which encode the longitudinal
and Hall conductivities, $\sigma$ and $\sigma_{\Hrm}$. To write
down the conductivities we will use the explicit plasmon propagator
\begin{align}
\dfrac{1}{\omega^{2}-\tilde{c}_{s}^{2}(\qv^{2})\qv^{2}} & =\dfrac{1}{\omega^{2}-c_{s}^{2}\qv^{2}-\omega_{p}^{2}|\qv|^{2-\alpha}},\label{eq:73}
\end{align}
which is gapped in the case $\alpha=2$, but has a gapless dispersion relation
$\omega\sim |\qv|^{1-\alpha/2}$ for $0\leq\alpha<2$. We then find the longitudinal and Hall [\cref{conSCgen}] conductivities,
\begin{align}
\sigma(\omega,\qv)={} & \dfrac{n_{0}}{m}\frac{\ir \omega}{\omega^{2}-|\qv|^{2-\alpha}\omega_{p}^{2}-c_{s}^{2}\qv^{2}},&
\sigma_{\Hrm}(\omega,\qv)={} & \frac{sn_{0}}{2m^{2}}\dfrac{-\qv^{2}}{\omega^{2}-|\qv|^{2-\alpha}\omega_{p}^{2}-c_{s}^{2}\qv^{2}}. \label{Hallcon}
\end{align}
Note that here and below, the superfluid expressions can be obtained
by setting $\omega_{p}=0$. We can also extract the density-density
response $\chi_{nn}=\ir q^{2}\sigma/\omega$ and verify using \cref{eq:69},
the compressibility sum-rule $K^{-1}(\qv)=n_{0}^{2}/\chi_{nn}(\omega=0,\qv)$,
which follows from the thermodynamic identity $K^{-1}=n^{2}\left(\delta\mu/\delta n\right)_{T}$.

The third group of terms in $\Wc$ includes those in which $\tilde{c}_{s}$
appears in both the numerator and the denominator. Since 
\begin{align}
\dfrac{\tilde{c}_{s}^{2}(\qv^{2})}{\omega^{2}-\tilde{c}_{s}^{2}(\qv^{2})\qv^{2}}=  \frac{1}{|\qv|^{\alpha}}\frac{c_{s}^{2}|\qv|^{\alpha}+\omega_{p}^{2}}{\omega^{2}-c_{s}^{2}\qv^{2}-\omega_{p}^{2}|\qv|^{2-\alpha}}
\sim & \begin{cases}
\frac{c_{s}^{2}+\omega_{p}^{2}}{\omega^{2}} &(\alpha=0)\\
\frac{1}{|\qv|^{\alpha}}\frac{\omega_{p}^{2}}{\omega^{2}} & (0<\alpha<2)\\
\frac{1}{\qv^{2}}\frac{\omega_{p}^{2}}{\omega^{2}-\omega_{p}^{2}} & (\alpha=2)
\end{cases}
\label{eq:75}
\end{align}
as $\qv\to 0$ at $\omega\neq0$, response functions of this
kind may diverge in this limit, for $0<\alpha\leq2$. On the other
hand, as is clear from the right hand side of \cref{eq:75}, response
functions in this group are identical in the superfluid and superconductor
at $\omega=0$ and $\qv\neq0$. A basic example is given by the term
$B^{2}$ included in $\mathsf{B}^{2}$ in the second line of \cref{eq:71},
which implies the London diamagnetic response, 
\begin{align}
\rho_{\Lrm}(\omega,\qv)= {}& -\frac{n_{0}}{m}\frac{1}{|\qv|^{\alpha}}\frac{c_{s}^{2}|\qv|^{\alpha}+\omega_{p}^{2}}{\omega^{2}-c_{s}^{2}\qv^{2}-\omega_{p}^{2}|\qv|^{2-\alpha}}.
\end{align}
Another term in this group is the $h^{2}$ term in the third line
of \cref{eq:71}, which implies that the``dynamic compressibility'' is given by
\begin{align}
K_{\dr}^{-1}(\omega,\qv)= {}& n_{0}m\frac{\omega^{2}}{|\qv|^{\alpha}}\frac{c_{s}^{2}|\qv|^{\alpha}+\omega_{p}^{2}}{\omega^{2}-c_{s}^{2}\qv^{2}-\omega_{p}^{2}|\qv|^{2-\alpha}}.
\end{align}
The dynamic compressibility is defined as the pressure response to volume changes, $K_{\text{d}}^{-1}=-4\delta^{2}\Wc/\delta h^{2}$,
and vanishes at $\omega=0$. On the other hand, $\lim_{\omega\rightarrow\infty}K_{\dr}^{-1}(\omega,\qv)=K^{-1}(\qv)$,
and $K_{\dr}^{-1}(\omega,\qv)\sim K^{-1}(\qv)\sim q^{-\alpha}$
as $q\to 0$.

The mixed responses $\kappa^{ij,k}=\delta T^{ij}/\delta E_{k}=2\delta J^{k}/\delta\partial_{t}h_{ij}$
and $\chi^{ij}=2\delta n/\delta h_{ij}$ are encoded in $E^{i}\partial_{i}h$ coming from $ hq_{i}\mathsf{E}^{i}$
in the third line of \labelcref{eq:71}, 
\begin{align}
\kappa^{ij,k}(\omega,\qv) & =n_{0}\delta^{ij}\ir q^{k}\frac{1}{|\qv|^{\alpha}}\frac{c_{s}^{2}|\qv|^{\alpha}+\omega_{p}^{2}}{\omega^{2}-c_{s}^{2}\qv^{2}-|\qv|^{2-\alpha}\omega_{p}^{2}},\\
\chi^{ij}(\omega,\qv) & =-n_{0}\delta^{ij}|\qv|^{2-\alpha}\frac{c_{s}^{2}|\qv|^{\alpha}+\omega_{p}^{2}}{\omega^{2}-c_{s}^{2}\qv^{2}-|\qv|^{2-\alpha}\omega_{p}^{2}},\label{eq:19-1}
\end{align}
 Finally, the second odd viscosity comes from $hq_{i}(\partial_t \omega^i - \partial^i \omega_t)$ contained in $hq_{i}\mathsf{E}^{i}$ in the induced action,
and is given by [\cref{EFT:eta12csSCmixed}]
\begin{align}
\eta_{\text{o}}^{(2)}(\omega,\qv^{2})= - \frac{1}{2} sn_{0} \dfrac{1}{|\qv|^{\alpha}}\frac{c_{s}^{2}|\qv|^{\alpha}+\omega_{p}^{2}}{\omega^{2}-c_{s}^{2}|\qv|^{2}-|\qv|^{2-\alpha}\omega_{p}^{2}}.\label{Hallvis}
\end{align}

\section{Details of microscopic calculation}
\label{app:micCalcs}

In this Appendix we provide details for the computations outlined in \cref{sec:microTheory}.

\subsection{Flat space induced action}
\label{app:MicCalcFSnoC}

In flat space ($h_{ij}=0$), we rewrite the action \labelcref{eq:effactForm} for the Nambu spinors in Fourier space as 
\begin{equation}
S[\Psiv^\dagger,\Psiv,\Ac] = -\dfrac{1}{2} \int_{q} \Psiv_q^\dagger \Gc^{-1}_{0,q} \Psiv_q + { \dfrac{1}{2}}\int_{q,q'}  \Psiv_q^\dagger [\Gamma_{1,q,-q'}+\Gamma_{2,q,-q'}] \Psiv_{q'}
\end{equation}
where the mean field propagator is given by \cref{eq:MFpropag} 
and we have split the vertex $\Gamma = \Gc_0^{-1}-\Gc^{-1}$ introduced in \cref{eq:defGammaGeneral} 
into two vertices $\Gamma_{1,2}$ defined by
\begin{align}
\Gamma_{1,q,-q'}&=-\ir \Ac_{0,q-q'} \sigma_z  - \dfrac{1}{2m} (\qv + \qv')\cdot \Avc_{q-q'} \sigma_0,&
\Gamma_{2,q,-q'}&=\dfrac{1}{2m} \int_p \Avc_{p}\cdot \Avc_{q-q'-p} \sigma_z. 
\end{align}
In terms of these vertices, the contributions to \cref{eq:loopExpansionSeff} to first and second order in  $\Ac$ are
\begin{gather}
S[\Ac] =   { \dfrac{1}{2}}\Tr [\Gc_0 \Gamma_{1}] + { \dfrac{1}{2}}\Tr [\Gc_0\Gamma_{2}] + { \dfrac{1}{4}}\Tr [\Gc_0 \Gamma_1 \Gc_0 \Gamma_1].
\end{gather}
The first two sums yield
\begin{align}
- { \dfrac{1}{2}}\Tr \Gc_0 \Gamma_{1} &= -\ir  \Ac_{0,q=0}n_0,&
- { \dfrac{1}{2}}\Tr \Gc_0 \Gamma_{2} &= \dfrac{n_0 }{2m} \int_p \Avc_{p}\cdot \Avc_{-p}, \label{eq:flatSpaceTrivialLoops}
\end{align}
with $n_0={ \frac{1}{2}}\braket{\Psiv^\dagger\sigma_z\Psiv} =  \frac{1}{2}\Tr [ \Gc_0\sigma_z]$ the density of electrons in the ground state in absence of external sources. 
The first term does not contribute to the response functions as it is linear in $A_\mu$ and $\partial_\mu \theta$ vanishes when evaluated at $q=0$.  

To evaluate the last integral, we use that with the shorthand notations $\kappa^\mu$ [\cref{eq:defQbubble}],
\begin{align}
\Gamma_{1,p+q,-q}= -(
\ir \Ac_{0,p}\kappa^0_{p,q} + \Ac_{j,p}\kappa^j_{p,q})
\end{align}
 hence
\begin{align}
{ \dfrac{1}{4}}\Tr [\Gc_0 \Gamma_1 \Gc_0 \Gamma_1] ={}&{} \dfrac{1}{2}\int_q\bigg\{ \Ac_{i,-q}\Ac_{j,q} \int_p { \dfrac{1}{2}} \tr [\kappa^i_{p,q} \Gc_{0,p}\kappa^j_{p,q} \Gc_{0,p+q}]\nnl
{}&{}+(\ir \Ac_{0,-q})\Ac_{j,q} \int_p { \dfrac{1}{2}} \tr [\kappa^0_{p,q} \Gc_{0,p}\kappa^j_{p,q} \Gc_{0,p+q}]
\nnl
{}&{}+\Ac_{i,-q}(\ir \Ac_{0,q}) \int_p { \dfrac{1}{2}} \tr [\kappa^i_{p,q} \Gc_{0,p}\kappa^0_{p,q} \Gc_{0,p+q}]\nnl
{}&{}+(\ir \Ac_{-q,0})( \ir \Ac_{q,0}) \int_p { \dfrac{1}{2}} \tr [\kappa^0_{p,q} \Gc_{0,p}\kappa^0_{p,q} \Gc_{0,p+q}]\bigg\}. \label{eq:App_action_gauge_phase_complex}
\end{align}
Introducing the correlators $Q^{\mu\nu}(q)$ [\cref{eq:defQbubble}] and transforming back to real time gives \cref{eq:action_gauge_phase_simpl}.
Integrating out the phase mode $\theta$ yields \cref{eq:action_gauge_phase_K}.

\subsection{Conductivity from the loop integrals}
\label{app:CalcQintegrals}

The correlations functions $Q^{\mu\nu}(q)$ are expressed in terms of the normal and anomalous propagators $G_q$ and $F_q$ ,
\begin{align}
Q^{00}(q) &=   \int_p G_{p}G_{p+q}-\re [F_{p}^*F_{p+q}],\\
Q^{i0}(q) &=\dfrac{1}{{ 2}m}\int_{p} (q^i+2p^i) \{G_{p}G_{p+q}+\ir \im [F_{p}^*F_{p+q}]\},\\
Q^{ij}(q) &=  {\dfrac{n_0}{m} \delta^{ij}}+\dfrac{1}{{ 4}m} \int_p(q^i+2p^i)(q^j+2p^j)\{ G_{p}G_{p+q}+\re [F_{p}^*F_{p+q}]\}.
\end{align}
These expressions can be further simplified by exploiting the space rotation symmetry of the theory and noticing that the functions are either odd or even under $\omega_n \to - \omega_n$. In particular, $Q^{i0}(\ir \omega_n,\qv)$ can be split into an even and an odd part,
\begin{gather}
Q^{i0}(\ir \omega_n,\qv)= Q^{i0}_\er(\ir \omega_n,\qv)+Q^{i\tau}_\orm(\ir \omega_n,\qv),\\
\begin{aligned}
Q_\er^{i0}(q)&=\dfrac{1}{{ 2}m}\int_{p}(q^i+2p^i) G_{p}G_{p+q},&
Q_\orm^{i0}(q)&= \dfrac{\ir }{{ 2}m}\int_{p}(q^i+2p^i) \im F_{p}^*F_{p+q}.
\end{aligned}
\label{app:eqDefQevenodd}
\end{gather}
While both $Q^{i0}_\er$ and $Q^{i0}_\orm$ transform like vectors, the latter is odd under both time-reversal and parity. The decomposition for all $Q^{\mu\nu}(\ir\omega_n,\qv)$ reads 
\begin{gather}
Q^{00}(\ir \omega_n,\qv) = Q^\tau(\omega_n^2,\qv^2),
\\
\begin{aligned}
Q^{i0}_\er(\ir \omega_n,\qv)&= -\ir \omega_n q^i Q^{\er}(\omega_n^2,\qv^2),&
Q^{i0}_\orm(\ir \omega_n,\qv) &= \ir \epsilon^{ik} q_k Q^{\orm}(\omega_n^2,\qv^2),
\end{aligned}
\\
Q^{ij}(\ir \omega_n,\qv) = \delta^{ij}Q^{A}( \omega_n^2,\qv^2)+q^i q^j Q^{B}( \omega_n^2,\qv^2).
\end{gather}
Each of the five scalar functions $Q^\tau$, $Q^{\er}$, $Q^{\orm}$, $Q^{A}$, $Q^{B}$ can be obtained by projecting the  $Q^{\mu\nu}$ onto the corresponding tensors.

This decomposition allows us to write the projection of the conductivity tensor onto its antisymmetric part as
\begin{align}
\sigma^\Hrm(\omega,\qv)=\dfrac{\epsilon_{ij}}{2}\dfrac{1}{\ir\omega^+}K^{\Rrm,ij}(\omega,\qv)
=
\dfrac{\qv^2Q^{\orm} (\qv^2 Q^{B}+Q^{A}+ \omega
   ^2 Q^{\er} )}{\qv^4 Q^{B}+\qv^2 Q^{A}+\omega ^2 (2 \qv^2
   Q^{\er}+Q^\tau)} \label{eq:fullSigHall}
\end{align}
with the functions on the rhs evaluated at the momentum $\qv$ and analytically continued to the real frequency $\omega$. The polarisation bubbles are regular at small $\qv$ and $\omega$ and in that limit \cref{eq:fullSigHall} reduces to 
\begin{equation} \label{eq:Hallapp}
\sigma^\Hrm(\omega\to 0,\qv\to 0)=Q^{\orm}(0,0) \dfrac{\qv^2 Q^{A}(0,0) }{\qv^2 Q^{A}(0,0) + \omega^2 Q^{\tau}(0,0)}
\end{equation}
up to order $\omega^2$, $\qv^2$.
These loop integrals evaluate to
\begin{align}
Q^{\tau}(0,0)&{}=-\dfrac{m}{2\pi},&
Q^{A}(0,0)&{}=\dfrac{n_0}{m},&
Q^{\orm}(0,0)&{}=\pm\dfrac{1}{8\pi}.
\end{align}
Substituting these results into \cref{eq:Hallapp} gives rise to  \cref{eq:exprSigHMicroNoC}.

\subsection{Pure geometric induced action}
\label{app:MicCalcCSnoC}

In this Appendix, we determine the contributions to the odd viscosities coming from $S_0[h]$. In the first step we compute $S_0[h]$ to quadratic order in $h_{ij}=g_{ij}-\delta_{ij}$. To that end, we expand $\Gamma=\Gc_0^{-1}-\Gc^{-1}$, defined by \cref{eq:defGammaGeneral}, in powers of $h$, setting $\Ac=0$.

First, we recall the expansions
\begin{align}
\sqrt{g} &= 1 + \dfrac{1}{2}\delta^{ij}h_{ij}+\dfrac{1}{2}h-\dfrac{1}{8}\delta^{ij}\delta^{kl}h_{ij}h_{kl}+\Oc(h^3),\\
e^{ia}&=\delta^{ia}-\dfrac{1}{2}h^{ia}+ { \dfrac{ 3}{8} }h^{ik} h_k{}^a+\Oc(h^3).
\end{align}
In the above expressions, we denote $h=\det(h_{ij})=\epsilon^{ik}\epsilon^{jl}h_{ij}h_{kl}$ and the indices of $h_{ij}$ (and $\delta_{ij}$) are raised and lowered using the flat metric, as well as the second index of the vielbeins. Using this, one has
\begin{align}
\Gamma_{x,x'}={}&{}
-\bigg(\dfrac{1}{2}\delta^{ij}h_{ij}+\dfrac{1}{2}h-\dfrac{1}{8}\delta^{ij}\delta^{kl}h_{ij}h_{kl}\bigg) \Gc_0^{-1} -\bigg(1+\dfrac{1}{2}\delta^{ij}h_{ij}\bigg)\nnl
{}&{} \times\bigg\{ \dfrac{1}{4}\partial_\tau \bigg(\delta^{ij}h_{ij} +  h - \dfrac{1}{2}\delta^{ij}\delta^{kl}h_{ij}h_{kl}\bigg)\nnl
{}&{}-\dfrac{h_{ij}p^{i}p^{j}}{2m} + \Delta (\dfrac{1}{2}h_{ia} -  \dfrac{ 1}{8} h_{ik} h^k{}_a) p^i \sigt^a
\bigg\}
\delta(x-x')
+\Oc(h^3).
\end{align}
There are two contributions of quadratic order in $h_{ij}$ to the induced action: $\Tr \Gamma^{(2)} \Gc_0$ and $\Tr \Gamma^{(1)} \Gc_0\Gamma^{(1)} \Gc_0$, with $\Gamma^{(i)}$ the term of $i$-th order in $h_{ij}$ in $\Gamma$. A first remark is that the tadpole term $\Tr \Gamma^{(2)} \Gc_0$ does not contribute to the anti-symmetric part of the viscosity.
To see it, one notes that $\Gamma^{(2)}$ can be expressed as
\begin{equation}
\Gamma^{(2)}_{x,x'}= [h_{ij}h_{kl} D_1^{ij,kl} + \partial_\tau (h_{ij}h_{kl}) D_2^{ij,kl}] \delta(x-x')
\end{equation}
with $D_{1,2}^{ij,kl}$ differential operators satisfying $D_{1,2}^{ij,kl}=D_{1,2}^{kl,ij}$. The corresponding induced action is $\propto \sum_p h_{-p,ij}h_{p,kl}\int_q\Gc_0(q) D_{1}^{ij,kl}(q)$ which is symmetric under the exchange of the two pairs of indices $(ij)$ and $(kl)$.

It is thus sufficient to expand $\Gamma$ to order one in $h_{ij}$ and, in Fourier space,
\begin{align}
\Gamma_{q,-q'}^{(1)} {}&{}= h_{ij}(q-q') \bigg\{\dfrac{1}{2}  \delta^{ij}
\bigg(
q'_0 \sigma^0 - \dfrac{q_k q'{}^l}{2m} \sigma^z + \Delta q'_k \sigt^k
\bigg) {} \nnl 
 {}&+{} \bigg[ 
\dfrac{1}{4} (q-q')_0 \delta^{ij} \sigma^0 - \dfrac{q^i q'{}^j + q'{}^i q^j}{4m} \sigma^z + \dfrac{1}{4} \Delta [q'{}^i \sigt^j + \sigt^i q'{}^j] 
\bigg]
\bigg\},
\end{align}
which yields the induced action $S_0 = \Tr [\Gamma^{(1)} \Gc_0\Gamma^{(1)} \Gc_0]/4$, see \labelcref{eq:KernelRac}.

\subsection{Pure geometric contribution to the odd viscosities}
\label{app:MicCalcEtaGeonoC}

The odd viscosities are obtained by projecting $R^{ij,kl}$ onto the odd tensors $\sigma^{ab}$, using the identity $(\sigma^{ab})^{ijkl}(\sigma^{cd})^{jilk}/8=\delta_{a,c}\delta_{b,d}-\delta_{a,d}\delta_{b,c}$. All loop integrals obtained while computing the trace are regular at small $\qv$, $\omega$ and the contribution to $\eta^{(2)}_\orm$ is of order $\Oc(|\qv|,\omega)$.
As for $\eta_{\orm}^{(1)}$, one has in terms of the propagators \labelcref{eq:defGF,eq:defsmallf}
\begin{align}
 \eta_{\orm}^{(1)}(\omega,\qv) &= \dfrac{1}{\ir \omega} \dfrac{\ir s \Delta}{16 m} \int_p
 \pv (\pv + \qv)  \{ [\pv (2\pv + \qv)f_p + m \Delta G_p ] G_{-p-q}\nnl
{}&{} -[\pv (2\pv + \qv)f_p + m \Delta G_{-p} ] G_{p+q}\}.
\label{eq:etaodd01Fullexpr}
\end{align}
To compute this expressions, we first recall that the fermion density is given by
\begin{align}
n_0={}&{}\dfrac{1}{2}\Tr[\Gc_0 \sigma_z]=-\dfrac{1}{2}\int_{\qv,\omega_n}\dfrac{(\ir \omega_n + \xi_\qv) \er^{\ir \omega_n \nu}+(-\ir \omega_n + \xi_\qv) \er^{-\ir \omega_n \nu}}{\omega_n^2+\qv^2\Delta^2+\xi_\qv^2}\nnl
={}&{}\dfrac{1}{2}\int_{\qv}1-\dfrac{\xi_\qv}{\sqrt{\qv^2\Delta^2+\xi_\qv^2}}=\int_{\qv}\dfrac{\qv^2\Delta^2(\qv^2/m-\xi_\qv)}{4(\qv^2\Delta^2+\xi_\qv^2)^{3/2}}.
\end{align}
We have introduced a convergence factor $\nu \to 0^+$ to take care of the time ordering of the operators. The last integral is obtained by performing an integration by parts $\sum_\qv f(\qv^2)=-\sum_\qv \qv^2 f'(\qv^2)$, discarding the boundary term.
Evaluating \cref{eq:etaodd01Fullexpr} at $q=0$ and carrying the integral over Matsubara frequencies, one gets
\begin{equation}
 \eta_{\orm}^{(1)}=s\int_{\qv}\dfrac{\qv^2\Delta^2(\qv^2/m-\xi_\qv)}{8(\qv^2\Delta^2+\xi_\qv^2)^{3/2}}=\dfrac{s n_0}{2}
\end{equation}
which is \cref{eq:exprEta1MicroNoC} in the main text.

\subsection{Phase term contribution to the odd viscosities}
\label{app:MicCalcPhaseContribnoC}

In this Appendix, we determine the contribution to the odd viscosities from the phase mode part of the induced action. To that end, we determine first $N^\mu[h]$, the coefficient of the term linear in $\Ac_\mu$
in the induced action $S[\Ac,h]$. We expand $\Gamma$, defined by \cref{eq:defGammaGeneral} in powers of $\Ac$.\footnote{Anticipating the analytic continuation we replace $\ir \Ac_\tau$ by $\Ac_\tau$.}  Denoting $\Gamma^{(i)}$ the term of order $\Oc(\Ac^i)$,\footnote{Note that this convention is different to that of \cref{app:MicCalcCSnoC} where $\Ac$ is dropped and $\Gamma^{(i)}$ denotes the terms of order $\Oc(h^i)$.}
we have
\begin{gather}
\Gamma^{(0)}_{p,-p'} = \Delta_0 p_i' (e^{ia}_{p-p'} - \delta^{ia} \delta_{p,p'})\sigt_a  + \dfrac{p_i p'_j+p_i' p_j}{4m} (g^{ij}_{p-p'} - \delta^{ij} \delta_{p,p'})\sigma_z  ,\\
 \begin{aligned}
\Gamma^{(1)}_{p,-p'}&=\int_q \Ac_{-q,\mu} \gamma^{(1)\mu}_{p,-p',q}, &
\gamma^{(1)0}_{p,-p',q} &=- e \delta_{p-p'+q} \sigma^z, &
\gamma^{(1)i}_{p,-p',q} &=-\dfrac{e}{2m}(p_j+p_j') g^{ij}_{p-p'+q} \sigma^0.
\end{aligned} 
\end{gather}
By identifying the induced action \labelcref{eq:loopExpansionSeff} with the expansion \labelcref{eq:inducedActionFormal}, we find
\begin{equation}
 N_{q}^\mu[h]= \dfrac{1}{2}\tr\int_p \Gc_{0,p} \gamma^{(1)\mu}_{p,-p,q}  + \dfrac{1}{2} \int_{p} \tr \Gc_{0,p} \gamma^{(1)}_{p,-p-q,q} \Gc_{0,p+q} \Gamma^{(0)}_{p+q,-p}.
\end{equation}
Computing $N_{q}^\mu$ at order $\Oc(h)$ is done by evaluating the bubble diagram $\Tr \Gc_0 \Gamma^{(1)} \Gc_0 \Gamma^{(0)}$, setting $h_{ij}=0$ in $\Gamma^{(1)}$ and keeping $ \Gamma^{(0)}$ to linear order in $h_{ij}$, which produces \cref{eq:defIqN,eq:exprIqN}.

To project the viscosity tensor deduced from \cref{eq:relEtaS,eq:defSijkl} onto the antisymetric tensors $\sigma^{ab}$, we expand $I^{ij}_q$ in the limit of small momentum at order $\Oc(\qv^2)$. $I^{ij}_q$ can be decomposed onto symmetric rank-two tensors that transform like $\qv$ under $\SO(2)$ rotations,\footnote{The term $u^{i}u^{j}(\qv\cdot \uv^*)^2$ as well as its complex conjugate is allowed by symmetry but does not appear in the calculation.}
\begin{align}
I^{ij}_q ={}&{} [I_{\delta}(\ir \omega_n)+\qv^2 I_{\delta,1}(\ir \omega_n)]\delta^{ij} + [I_{\uv\uv}(\ir \omega_n)+\qv^2 I_{\uv\uv,1}(\ir \omega_n)] u^{(i}u^{*j)}\nnl
{}&{}  +I_{\qv \uv}(\ir \omega_n) q^{(i}u^{j)}(\qv \cdot \uv^*) +I_{\qv \uv^*}(\ir \omega_n) q^{(i}u^{*j)}(\qv \cdot \uv^*) + I_{\qv \qv}(\ir \omega_n)q^i q^j +\Oc(\qv^2)
\label{eq:defIalp}
\end{align}
where the integrals $I_{\alpha}$ are functions of the external frequency $\ir \omega_n$. 

No matter what are the values of the integrals  $I_{\alpha}$, the projection of $S^{ijkl}$ onto $\sigma^{xz}$ vanishes and there is no contribution to $\eta^{(1)}_\orm$.
 We now extract the viscosity $\eta^{(2)}_\orm$ from the integrals $I_\alpha$ defined by \cref{eq:defIalp}.
 Expanding at small frequencies and noting $I_{\alpha}(\ir \omega_n) = I_{\alpha}(0) + \ir \omega_n I_{\alpha}'(0)+\Oc(\omega_n^2)$, the projection yields
\begin{equation}
\eta_\orm^{(2)}(\qv,\omega)=  8 s \dfrac{I_{\qv \uv}(0) [I_{\delta}'(0) + I_{\uv \uv}'(0)]}{n/m(c_s^2\omega^2-\qv^2)}.
\end{equation}
To get this result, we have used that $I_{\qv \uv}(0) = - I_{\qv \uv^*}(0)$,  $I_{\qv \uv}'(0) = I_{\qv \uv^*}'(0)$, and as shown in \cref{app:CalcQintegrals} $q_\alpha q_\beta Q^{\alpha \beta}(q)=-n/m(c_s^{-2}\omega^2-\qv^2)$ at small frequency and momentum.
The relevant integrals are
\begin{align}
I_{\delta}(\ir \omega_n) =& -\dfrac{\ir  \omega_n}{8 m} \int_{\pv,\ir \omega_m} \pv^2 [\pv^2
   f(\pv,\ir \omega_m) f(\pv,\ir \omega_m)-G(-\pv, -\ir \omega_m) G(-\pv,-\ir \omega_m) \nnl
   &{} -G(\pv,\ir \omega_m)
   G(\pv,\ir \omega_m)]
+\Oc(\omega_n^2),
   \\
I_{\uv \uv}(\ir \omega_n) =& \frac{\ir \omega_n }{8 m}  \int_{\pv,\ir \omega_m} \pv^2 \{f(\pv,\ir \omega_m)
   [2 \Delta  m (G(-\pv,-\ir \omega_m)+G(\pv,\ir \omega_m))-\pv^2
   f(\pv,\ir \omega_m)]\nnl
   &{}+2 \Delta  m (G(-\pv,-\ir \omega_m)+G(\pv,\ir \omega_m))
   f(\pv,\ir \omega_m)\} 
+\Oc(\omega_n^2), \\
I_{\qv \uv}(\ir \omega_n) =&\frac{4 \Delta  m}{32 m^2} \int_{\pv,\ir \omega_m} \pv^2 G(-\pv,-\ir \omega_m)
   [f(\pv,\ir \omega_m)+\pv^2
   f'(\pv,\ir \omega_m)]\nnl
   &{}-|\pv|^4 f(\pv,\ir \omega_m)
   [f(\pv,\ir \omega_m)+4 \Delta  m
   G'(\pv,\ir \omega_m)]
   +\Oc(\omega_n),
\end{align}
where $G'=\partial_{\qv^2} G$, $f' =\partial_{\qv^2} f$.
As in \cref{app:MicCalcCSnoC}, we perform for each integral the sum over Matsubara frequencies and  obtain ultraviolet-divergent integrals. We regularize them using a hard momentum cutoff and compare these  integrals to those defining the density by identifying the leading logarithmic term. We thus obtain $I_{\qv \uv}(0) = n_0/32m$, $I_{\delta}'(0) + I_{\uv \uv}'(0) = -2 n_0$ to finally get \cref{eq:exprEta2MicroNoC}.

\end{appendix}


\begin{thebibliography}{10}
\providecommand{\url}[1]{\texttt{#1}}
\providecommand{\urlprefix}{URL }
\expandafter\ifx\csname urlstyle\endcsname\relax
  \providecommand{\doi}[1]{doi:\discretionary{}{}{}#1}\else
  \providecommand{\doi}{doi:\discretionary{}{}{}\begingroup
  \urlstyle{rm}\Url}\fi
\providecommand{\eprint}[2][]{\url{#2}}

\bibitem{Read2000}
N.~{Read} and D.~{Green},
\newblock \emph{{Paired states of fermions in two dimensions with breaking of
  parity and time-reversal symmetries and the fractional quantum Hall effect}},
\newblock Phys. Rev. B \textbf{61}, 10267 (2000),
\newblock \doi{10.1103/PhysRevB.61.10267}.

\bibitem{Alicea2012}
J.~{Alicea},
\newblock \emph{{New directions in the pursuit of Majorana fermions in solid
  state systems}},
\newblock Rep. Prog. Phys. \textbf{75}(7), 076501 (2012),
\newblock \doi{10.1088/0034-4885/75/7/076501}.

\bibitem{volovikbook}
G.~E. Volovik,
\newblock \emph{{The Universe in a Helium Droplet}}, vol. 117,
\newblock Oxford University Press, New York (2009).

\bibitem{Levitin2013}
L.~V. Levitin, R.~G. Bennett, A.~Casey, B.~Cowan, J.~Saunders, D.~Drung,
  T.~Schurig and J.~M. Parpia,
\newblock \emph{{Phase Diagram of the Topological Superfluid $^3\mathrm{He}$
  Confined in a Nanoscale Slab Geometry}},
\newblock Science \textbf{340}(6134), 841 (2013),
\newblock \doi{10.1126/science.1233621}.

\bibitem{Shook2019}
A.~Shook, V.~Vadakkumbatt, P.~Senarath~Yapa, C.~Doolin, R.~Boyack, P.~Kim,
  G.~Popowich, F.~Souris, H.~Christani, J.~Maciejko and et~al.,
\newblock \emph{{Stabilized Pair Density Wave via Nanoscale Confinement of
  Superfluid $^3\mathrm{He}$}},
\newblock Phys. Rev. Lett. \textbf{124}(1), 015301 (2020),
\newblock \doi{10.1103/physrevlett.124.015301}.

\bibitem{kallin2016chiral}
C.~Kallin and J.~Berlinsky,
\newblock \emph{Chiral superconductors},
\newblock Rep. Prog. Phys. \textbf{79}(5), 054502 (2016),
\newblock \doi{10.1088/0034-4885/79/5/054502}.

\bibitem{Xu2015}
J.-P. Xu, M.-X. Wang, Z.~L. Liu, J.-F. Ge, X.~Yang, C.~Liu, Z.~A. Xu, D.~Guan,
  C.~L. Gao, D.~Qian, Y.~Liu, Q.-H. Wang \emph{et~al.},
\newblock \emph{{Experimental Detection of a Majorana Mode in the core of a
  Magnetic Vortex inside a Topological Insulator-Superconductor
  ${\mathrm{Bi}}_{2}{\mathrm{Te}}_{3}/{\mathrm{NbSe}}_{2}$ Heterostructure}},
\newblock Phys. Rev. Lett. \textbf{114}, 017001 (2015),
\newblock \doi{10.1103/PhysRevLett.114.017001}.

\bibitem{Menard2017}
G.~C. M\'{e}nard, S.~Guissart, C.~Brun, R.~T. Leriche, M.~Trif,
  F.~Debontridder, D.~Demaille, D.~Roditchev, P.~Simon and T.~Cren,
\newblock \emph{{Two-dimensional topological superconductivity in
  $\mathrm{Pb/Co/Si(111)}$}},
\newblock Nat. Commun. \textbf{8}(1) (2017),
\newblock \doi{10.1038/s41467-017-02192-x}.

\bibitem{banerjee2018observation}
M.~Banerjee, M.~Heiblum, V.~Umansky, D.~E. Feldman, Y.~Oreg and A.~Stern,
\newblock \emph{{Observation of half-integer thermal Hall conductance}},
\newblock Nature \textbf{559}(7713), 205–210 (2018),
\newblock \doi{10.1038/s41586-018-0184-1}.

\bibitem{goryo1998abelian}
J.~Goryo and K.~Ishikawa,
\newblock \emph{{Abelian Chern-Simons term in superfluid
  $\operatorname{^3He-A}$}},
\newblock Phys. Lett. A \textbf{246}(6), 549–559 (1998),
\newblock \doi{10.1016/s0375-9601(98)00438-1}.

\bibitem{horovitz2002spontaneous}
B.~Horovitz and A.~Golub,
\newblock \emph{{Spontaneous magnetization and Hall effect in superconductors
  with broken time-reversal symmetry}},
\newblock Europhys. Lett. \textbf{57}(6), 892–897 (2002),
\newblock \doi{10.1209/epl/i2002-00594-y}.

\bibitem{PhysRevB.77.174513}
R.~Roy and C.~Kallin,
\newblock \emph{{Collective modes and electromagnetic response of a chiral
  superconductor}},
\newblock Phys. Rev. B \textbf{77}, 174513 (2008),
\newblock \doi{10.1103/PhysRevB.77.174513}.

\bibitem{Lutchyn2008}
R.~M. Lutchyn, P.~Nagornykh and V.~M. Yakovenko,
\newblock \emph{{Gauge-invariant electromagnetic response of a chiral
  $p_x+ip_y$ superconductor}},
\newblock Phys. Rev. B \textbf{77}(14) (2008),
\newblock \doi{10.1103/physrevb.77.144516}.

\bibitem{Hoyos2013}
C.~Hoyos, S.~Moroz and D.~T. Son,
\newblock \emph{{Effective theory of chiral two-dimensional superfluids}},
\newblock Phys. Rev. B \textbf{89}, 174507 (2013),
\newblock \doi{10.1103/PhysRevB.89.174507}.

\bibitem{Avron1995}
J.~E. Avron, R.~Seiler and P.~G. Zograf,
\newblock \emph{{Viscosity of Quantum Hall fluids}},
\newblock Phys. Rev. Lett. \textbf{75}(4) (1995),
\newblock \doi{10.1103/PhysRevLett.75.697}.

\bibitem{Avron1997}
J.~E. Avron,
\newblock \emph{{Odd Viscosity}},
\newblock J. Stat. Phys. \textbf{92}(3-4), 543 (1998),
\newblock \doi{10.1023/A:1023084404080}.

\bibitem{Hoyos2014}
C.~Hoyos,
\newblock \emph{{Hall viscosity, topological states and effective theories}},
\newblock Int. J. Mod. Phys. B \textbf{28}(15), 1430007 (2014),
\newblock \doi{10.1142/s0217979214300072}.

\bibitem{PhysRevB.94.125427}
M.~Sherafati, A.~Principi and G.~Vignale,
\newblock \emph{Hall viscosity and electromagnetic response of electrons in
  graphene},
\newblock Phys. Rev. B \textbf{94}, 125427 (2016),
\newblock \doi{10.1103/PhysRevB.94.125427}.

\bibitem{PhysRevLett.119.226602}
L.~V. Delacr\'etaz and A.~Gromov,
\newblock \emph{{Transport Signatures of the Hall Viscosity}},
\newblock Phys. Rev. Lett. \textbf{119}, 226602 (2017),
\newblock \doi{10.1103/PhysRevLett.119.226602}.

\bibitem{PhysRevLett.118.226601}
T.~Scaffidi, N.~Nandi, B.~Schmidt, A.~P. Mackenzie and J.~E. Moore,
\newblock \emph{{Hydrodynamic Electron Flow and Hall Viscosity}},
\newblock Phys. Rev. Lett. \textbf{118}, 226601 (2017),
\newblock \doi{10.1103/PhysRevLett.118.226601}.

\bibitem{holder2019unified}
T.~Holder, R.~Queiroz and A.~Stern,
\newblock \emph{{Unified Description of the Classical Hall Viscosity}},
\newblock Phys. Rev. Lett. \textbf{123}, 106801 (2019),
\newblock \doi{10.1103/PhysRevLett.123.106801}.

\bibitem{PhysRevB.100.115421}
M.~Sherafati and G.~Vignale,
\newblock \emph{Hall viscosity and nonlocal conductivity of gapped graphene},
\newblock Phys. Rev. B \textbf{100}, 115421 (2019),
\newblock \doi{10.1103/PhysRevB.100.115421}.

\bibitem{PhysRevE.90.063005}
A.~Lucas and P.~Sur\'owka,
\newblock \emph{{Phenomenology of nonrelativistic parity-violating
  hydrodynamics in 2+1 dimensions}},
\newblock Phys. Rev. E \textbf{90}, 063005 (2014),
\newblock \doi{10.1103/PhysRevE.90.063005}.

\bibitem{PhysRevB.96.174524}
T.~I. Tuegel and T.~L. Hughes,
\newblock \emph{{Hall viscosity and the acoustic Faraday effect}},
\newblock Phys. Rev. B \textbf{96}, 174524 (2017),
\newblock \doi{10.1103/PhysRevB.96.174524}.

\bibitem{PhysRevFluids.2.094101}
S.~Ganeshan and A.~G. Abanov,
\newblock \emph{Odd viscosity in two-dimensional incompressible fluids},
\newblock Phys. Rev. Fluids \textbf{2}, 094101 (2017),
\newblock \doi{10.1103/PhysRevFluids.2.094101}.

\bibitem{banerjee2017odd}
D.~Banerjee, A.~Souslov, A.~G. Abanov and V.~Vitelli,
\newblock \emph{Odd viscosity in chiral active fluids},
\newblock Nat. Commun. \textbf{8}(1), 1573 (2017),
\newblock \doi{10.1038/s41467-017-01378-7}.

\bibitem{PhysRevLett.122.128001}
A.~Souslov, K.~Dasbiswas, M.~Fruchart, S.~Vaikuntanathan and V.~Vitelli,
\newblock \emph{{Topological Waves in Fluids with Odd Viscosity}},
\newblock Phys. Rev. Lett. \textbf{122}, 128001 (2019),
\newblock \doi{10.1103/PhysRevLett.122.128001}.

\bibitem{abanov2019hydrodynamics}
A.~G. Abanov, T.~Can, S.~Ganeshan and G.~M. Monteiro,
\newblock \emph{Hydrodynamics of two-dimensional compressible fluid with broken
  parity: variational principle and free surface dynamics in the absence of
  dissipation},
\newblock arXiv:1907.11196  (2019).

\bibitem{PhysRevLett.123.026804}
I.~S. Burmistrov, M.~Goldstein, M.~Kot, V.~D. Kurilovich and P.~D. Kurilovich,
\newblock \emph{{Dissipative and Hall Viscosity of a Disordered 2D Electron
  Gas}},
\newblock Phys. Rev. Lett. \textbf{123}, 026804 (2019),
\newblock \doi{10.1103/PhysRevLett.123.026804}.

\bibitem{narozhny2019magnetohydrodynamics}
B.~N. Narozhny and M.~Sch{\"u}tt,
\newblock \emph{{Magnetohydrodynamics in graphene: shear and Hall
  viscosities}},
\newblock Phys. Rev. B \textbf{100}(3), 035125 (2019),
\newblock \doi{10.1103/PhysRevB.100.035125}.

\bibitem{PhysRevB.100.115401}
S.~S. Apostolov, D.~A. Pesin and A.~Levchenko,
\newblock \emph{{Magnetodrag in the hydrodynamic regime: Effects of
  magnetoplasmon resonance and Hall viscosity}},
\newblock Phys. Rev. B \textbf{100}, 115401 (2019),
\newblock \doi{10.1103/PhysRevB.100.115401}.

\bibitem{soni2018free}
V.~Soni, E.~Bililign, S.~Magkiriadou, S.~Sacanna, D.~Bartolo, M.~J. Shelley and
  W.~Irvine,
\newblock \emph{{The odd free surface flows of a colloidal chiral fluid}},
\newblock Nat. Phys. \textbf{15}, 1188–1194 (2019),
\newblock \doi{10.1038/s41567-019-0603-8}.

\bibitem{Berdyugineaau0685}
A.~I. Berdyugin, S.~G. Xu, F.~M.~D. Pellegrino, R.~Krishna~Kumar, A.~Principi,
  I.~Torre, M.~Ben~Shalom, T.~Taniguchi, K.~Watanabe, I.~V. Grigorieva,
  M.~Polini, A.~K. Geim \emph{et~al.},
\newblock \emph{{Measuring Hall viscosity of graphene{\textquoteright}s
  electron fluid}},
\newblock Science \textbf{364}, 162 (2019),
\newblock \doi{10.1126/science.aau0685}.

\bibitem{Golan2019}
O.~Golan, C.~Hoyos and S.~Moroz,
\newblock \emph{Boundary central charge from bulk odd viscosity: Chiral
  superfluids},
\newblock Phys. Rev. B \textbf{100}, 104512 (2019),
\newblock \doi{10.1103/PhysRevB.100.104512}.

\bibitem{Read2009}
N.~Read,
\newblock \emph{{Non-Abelian adiabatic statistics and Hall viscosity in quantum
  Hall states and $p_x+ip_y$ paired superfluids}},
\newblock Phys. Rev. B \textbf{79}(4), 45308 (2009),
\newblock \doi{10.1103/PhysRevB.79.045308}.

\bibitem{Read2011}
N.~Read and E.~H. Rezayi,
\newblock \emph{{Hall viscosity, orbital spin, and geometry: Paired superfluids
  and quantum Hall systems}},
\newblock Phys. Rev. B \textbf{84}(8), 85316 (2011),
\newblock \doi{10.1103/PhysRevB.84.085316}.

\bibitem{taylor2010viscosity}
E.~Taylor and M.~Randeria,
\newblock \emph{{Viscosity of strongly interacting quantum fluids: spectral
  functions and sum rules}},
\newblock Phys. Rev. A \textbf{81}(5), 053610 (2010),
\newblock \doi{10.1103/PhysRevA.81.053610}.

\bibitem{Hoyos2012}
C.~Hoyos and D.~T. Son,
\newblock \emph{{Hall Viscosity and Electromagnetic Response}},
\newblock Phys. Rev. Lett. \textbf{108}, 066805 (2012),
\newblock \doi{10.1103/PhysRevLett.108.066805}.

\bibitem{Bradlyn2012}
B.~Bradlyn, M.~Goldstein and N.~Read,
\newblock \emph{{Kubo formulas for viscosity: Hall viscosity, Ward identities,
  and the relation with conductivity}},
\newblock Phys. Rev. B \textbf{86}(24), 245309 (2012),
\newblock \doi{10.1103/PhysRevB.86.245309}.

\bibitem{Geracie2015}
M.~Geracie, D.~T. Son, C.~Wu and S.-F. Wu,
\newblock \emph{{Spacetime symmetries of the quantum Hall effect}},
\newblock Phys. Rev. D \textbf{91}, 045030 (2015),
\newblock \doi{10.1103/PhysRevD.91.045030}.

\bibitem{Marino2018}
E.~C. Marino, D.~Niemeyer, V.~S. Alves, T.~H. Hansson and S.~Moroz,
\newblock \emph{{Screening and topological order in thin superconducting
  films}},
\newblock New J. Phys. \textbf{20}(8), 083049 (2018),
\newblock \doi{10.1088/1367-2630/aadb36}.

\bibitem{boyack2019electromagnetic}
R.~Boyack and P.~L. e.~S. Lopes,
\newblock \emph{Electromagnetic response of superconductors in the presence of
  multiple collective modes},
\newblock Phys. Rev. B \textbf{101}(9) (2020),
\newblock \doi{10.1103/physrevb.101.094509}.

\bibitem{Halperin1993}
B.~I. Halperin, P.~A. Lee and N.~Read,
\newblock \emph{{Theory of the half-filled Landau level}},
\newblock Phys. Rev. B \textbf{47}, 7312 (1993),
\newblock \doi{10.1103/PhysRevB.47.7312}.

\bibitem{Son2015}
D.~T. Son,
\newblock \emph{{Is the Composite Fermion a Dirac Particle?}},
\newblock Phys. Rev. X \textbf{5}, 031027 (2015),
\newblock \doi{10.1103/PhysRevX.5.031027}.

\bibitem{Mooij1990}
J.~E. Mooij, B.~J. van Wees, L.~J. Geerligs, M.~Peters, R.~Fazio and
  G.~Sch\"on,
\newblock \emph{Unbinding of charge-anticharge pairs in two-dimensional arrays
  of small tunnel junctions},
\newblock Phys. Rev. Lett. \textbf{65}, 645 (1990),
\newblock \doi{10.1103/PhysRevLett.65.645}.

\bibitem{Fazio1991}
R.~Fazio and G.~Sch\"on,
\newblock \emph{Charge and vortex dynamics in arrays of tunnel junctions},
\newblock Phys. Rev. B \textbf{43}, 5307 (1991),
\newblock \doi{10.1103/PhysRevB.43.5307}.

\bibitem{Diamantini1996}
M.~C. Diamantini, P.~Sodano and C.~Trugenberger,
\newblock \emph{{Gauge theories of Josephson junction arrays}},
\newblock Nucl. Phys. B \textbf{474}(3), 641 (1996),
\newblock \doi{10.1016/0550-3213(96)00309-4}.

\bibitem{Baturina2013}
T.~I. Baturina and V.~M. Vinokur,
\newblock \emph{Superinsulator–superconductor duality in two dimensions},
\newblock Ann. Phys. \textbf{331}, 236–257 (2013),
\newblock \doi{10.1016/j.aop.2012.12.007}.

\bibitem{Lucas2014}
A.~Lucas and P.~Surówka,
\newblock \emph{{Sound-induced vortex interactions in a zero-temperature
  two-dimensional superfluid}},
\newblock Phys. Rev. A \textbf{90}(5) (2014),
\newblock \doi{10.1103/physreva.90.053617}.

\bibitem{Hughes2013}
T.~L. Hughes, R.~G. Leigh and O.~Parrikar,
\newblock \emph{{Torsional anomalies, Hall viscosity, and bulk-boundary
  correspondence in topological states}},
\newblock Phys. Rev. D \textbf{88}(2), 025040 (2013),
\newblock \doi{10.1103/PhysRevD.88.025040}.

\bibitem{Barkeshli2012}
M.~Barkeshli, S.~B. Chung and X.-L. Qi,
\newblock \emph{{Dissipationless phonon Hall viscosity}},
\newblock Phys. Rev. B \textbf{85}, 245107 (2012),
\newblock \doi{10.1103/PhysRevB.85.245107}.

\bibitem{PhysRevLett.120.217002}
T.~Kvorning, T.~H. Hansson, A.~Quelle and C.~M. Smith,
\newblock \emph{{Proposed Spontaneous Generation of Magnetic Fields by Curved
  Layers of a Chiral Superconductor}},
\newblock Phys. Rev. Lett. \textbf{120}, 217002 (2018),
\newblock \doi{10.1103/PhysRevLett.120.217002}.

\bibitem{Son2006}
D.~T. Son and M.~Wingate,
\newblock \emph{{General coordinate invariance and conformal invariance in
  nonrelativistic physics: Unitary Fermi gas}},
\newblock Ann. Phys. \textbf{321}, 197 (2006),
\newblock \doi{10.1016/j.aop.2005.11.001}.

\bibitem{Greiter:1989qb}
M.~Greiter, F.~Wilczek and E.~Witten,
\newblock \emph{{Hydrodynamic Relations in Superconductivity}},
\newblock Mod. Phys. Lett. B \textbf{3}, 903 (1989),
\newblock \doi{10.1142/S0217984989001400}.

\bibitem{Moroz2014a}
S.~Moroz and C.~Hoyos,
\newblock \emph{{Effective theory of two-dimensional chiral superfluids: Gauge
  duality and Newton-Cartan formulation}},
\newblock Phys. Rev. B \textbf{91}, 064508 (2015),
\newblock \doi{10.1103/PhysRevB.91.064508}.

\bibitem{PhysRevB.93.024521}
S.~Moroz, C.~Hoyos and L.~Radzihovsky,
\newblock \emph{Chiral $p\ifmmode\pm\else\textpm\fi{}ip$ superfluid on a
  sphere},
\newblock Phys. Rev. B \textbf{93}, 024521 (2016),
\newblock \doi{10.1103/PhysRevB.93.024521}.

\bibitem{PhysRevB.94.125137}
A.~Quelle, C.~M. Smith, T.~Kvorning and T.~H. Hansson,
\newblock \emph{{Edge Majoranas on locally flat surfaces: The cone and the
  M\"obius band}},
\newblock Phys. Rev. B \textbf{94}, 125137 (2016),
\newblock \doi{10.1103/PhysRevB.94.125137}.

\bibitem{Golkar2014}
S.~Golkar, M.~M. Roberts and D.~T. Son,
\newblock \emph{{The Euler current and parity odd transport}},
\newblock J. High Energ. Phys. \textbf{04}, 110 (2015),
\newblock \doi{10.1007/JHEP04(2015)110}.

\bibitem{jiang2019geometric}
Q.-D. Jiang, T.~Hansson and F.~Wilczek,
\newblock \emph{{Geometric Induction in Chiral Superconductors}},
\newblock Phys. Rev. Lett. \textbf{124}, 197001 (2020),
\newblock \doi{10.1103/PhysRevLett.124.197001}.

\bibitem{souslov2019anisotropic}
A.~{Souslov}, A.~{Gromov} and V.~{Vitelli},
\newblock \emph{{Anisotropic odd viscosity via time-modulated drive}},
\newblock Phys. Rev. E \textbf{101}, 052606 (2020),
\newblock \doi{10.1103/PhysRevE.101.052606}.

\bibitem{rao2019hall}
P.~Rao and B.~Bradlyn,
\newblock \emph{Hall Viscosity in Quantum Systems with Discrete Symmetry: Point Group and Lattice Anisotropy},
\newblock Phys. Rev. X \textbf{10}, 021005 (2020),
\newblock \doi{10.1103/PhysRevX.10.021005}.

\bibitem{hawking1977zeta}
S.~W. Hawking,
\newblock \emph{{Zeta function regularization of path integrals in curved
  spacetime}},
\newblock Commun. Math. Phys. \textbf{55}(2), 133 (1977),
\newblock \doi{10.1007/BF01626516}.

\bibitem{PhysRevLett.42.1195}
K.~Fujikawa,
\newblock \emph{{Path-Integral Measure for Gauge-Invariant Fermion Theories}},
\newblock Phys. Rev. Lett. \textbf{42}, 1195 (1979),
\newblock \doi{10.1103/PhysRevLett.42.1195}.

\bibitem{PhysRevD.35.3796}
D.~J. Toms,
\newblock \emph{Functional measure for quantum field theory in curved
  spacetime},
\newblock Phys. Rev. D \textbf{35}, 3796 (1987),
\newblock \doi{10.1103/PhysRevD.35.3796}.

\end{thebibliography}
\end{document}